\documentclass[aps,preprint]{revtex4}%
\usepackage{amsfonts}
\usepackage{amsmath}
\usepackage{amssymb}
\usepackage{graphicx}%
\providecommand{\U}[1]{\protect\rule{.1in}{.1in}}

\begin{document}
\title[ ]{Equilibrium for Classical Zero-Point Radiation: Detailed Balance Under
Scattering by a Classical Charged Harmonic Oscillator}
\author{Timothy H. Boyer}
\affiliation{Department of Physics, City College of the City University of New York, New
York, New York 10031}

\begin{abstract}
It has been shown repeatedly over a period of 50 years that the use of
\textit{relativistic} classical physics and the inclusion of classical
electromagnetic zero-point radiation leads to the Planck blackbody spectrum
for classical radiation equilibrium. \ However, none of this work involves
scattering calculations. \ In contrast to this work, currently accepted
physical theory connects classical physics to only the Rayleigh-Jeans
spectrum. \ Indeed, in the past, it has been shown that a \textit{nonlinear}
classical oscillator (which is necessarily a \textit{nonrelativistic}
scattering system) achieves equilibrium only for the Rayleigh-Jeans spectrum
where the random radiation present at the frequency of the second harmonic of
the oscillator motion has the \textit{same} energy per normal mode as the
radiation present at the fundamental frequency. \ Here we continue work
emphasizing the importance of \ relativistic versus nonrelativistic analysis.
\ We consider the scattering of random classical radiation by a charged
harmonic oscillator of small but non-zero oscillatory amplitude (which can be
considered as a \textit{relativistic} scattering system) and show that
detailed radiation balance holds not only at the fundamental frequency of the
oscillator but through the first harmonic corresponding to quadrupole
scattering, provided that the radiation energy per normal mode at the first
harmonic is \textit{double} the radiation energy per normal mode at the
fundamental frequency. \ This condition corresponds exactly to the zero-point
radiation spectrum which is linear in frequency. \ It is suggested that for
this relativistic scattering system, the detailed balance for zero-point
radiation holds not only for the fundamental and first harmonic but extends to
all harmonics. \ Here we have the first example of an explicit
\textit{relativistic} classical \textit{scattering} calculation; equilibrium
corresponds not to the Rayleigh-Jeans spectrum, but rather corresponds to the
Lorentz-invariant zero-point radiation spectrum. \ 

\end{abstract}
\maketitle

\section{Introduction}

\subsubsection{Aspects Missing from the Classical Physics of 1900}

One suspects that the history of physics would be quite different if the
physicists at the turn of the 20th century had not been unaware of two crucial
ideas of classical physics: 1) the existence of classical electromagnetic
zero-point radiation, and 2) the importance of special relativity. \ Had the
earlier physicists included these aspects, they would have extended the
explanatory power of classical physics to include blackbody radiation, the
decrease of specific heats at low temperature, and the behavior of van der
Waals forces. \ As it happened, these aspects were first described in
connection with quantum ideas, and the use of quantum physics has been
extended until it is the dominant physical theory of the present day. \ Only
during the past half-century has a small group of physicists gone back to
classical physics to note the extensions of that theory made possible by the
inclusion of classical zero-point radiation and relativity.\cite{M1}%
\cite{B1975a}\cite{delaPena1996}\cite{Cole2003a}\cite{H-B2015}

The problem of the equilibrium spectrum of random radiation, the blackbody
radiation problem, provided one of the dilemmas for physics at the turn of the
20th century. \ The work on blackbody radiation within classical physics has
been reviewed recently.\cite{B2018c} \ Although the work involves a variety of
points of view, there is no treatment of the equilibrium spectrum of radiation
under \textit{scattering }by a relativistic classical system. \ Here we
provide the first \textit{relativistic scattering calculation} and show that
\textit{classical zero-point radiation} is indeed an equilibrium spectrum. \ 

Relativity imposes strong restrictions on systems. \ Two \textit{relativistic}
mechanical systems are 1) relativistic point charges in classical
electrodynamics, and 2) small harmonic oscillator systems within classical
electromagnetism. \ The first system (that of relativistic point charges in
classical electrodynamics) is a familiar relativistic system; the second
system involving harmonic oscillators is not obviously relativistic.
\ However, one can imagine a harmonic oscillator system of small amplitude as
arising from the classical electrodynamic system where two identical charged
particles $q$ are held at some fixed distance apart, and a third charged
particle $e$ of the same sign is placed between the two charges $q$ and is
allowed to oscillate along the line connecting the two identical charges $q$.
\ In the approximation of small amplitude of oscillation, the electrostatic
potential experienced by the third particle $e$ becomes a harmonic potential,
and the relativistic and nonrelativistic particle motions for the particle $e$
agree with each other since higher powers of the particle velocity can be
ignored. \ Within this simple harmonic motion of small amplitude (ignoring
terms in $v/c$), there is no role for the speed of light in vacuum $c$. \ \ 

\subsubsection{Radiation-Spectrum Stability}

If this harmonic-oscillator system involving the oscillating charge $e$ is
bathed in random classical electromagnetic radiation, then (using only the
dipole approximation for radiation) the harmonic oscillator system comes to
equilibrium at an oscillator energy equal to the energy of the radiation
normal modes of the same frequency as the oscillator frequency. \ This
energy-balance result connecting a point harmonic-oscillator system with
random radiation has been known since Planck's work at the end of the 19th
century.\cite{Planck} \ What does not seem to be recognized is that this
mechanical system, when treated beyond the dipole approximation, involves
\textit{detailed radiation balance at the radiation harmonics} of the
fundamental oscillator frequency provided that the random radiation spectrum
corresponds to that of classical electromagnetic \textit{zero-point}
radiation. \ 

Most physicists are satisfied to repeat the erroneous textbook claim\cite{err}
that classical physics leads inevitably to the Rayleigh-Jeans spectrum for
radiation equilibrium. \ The truth is far more nuanced.\cite{B2018c} \ It is
indeed true that if one use a \textit{nonrelativistic} classical theory such
as (nonrelativistic) classical statistical mechanics\cite{ER2p12} or considers
scattering by a \textit{nonrelativistic} \textit{nonlinear} dipole oscillator
treated in the dipole radiation approximation,\cite{B1976a} then one arrives
at the Rayleigh-Jeans spectrum. \ However, if one uses \textit{relativistic}
classical physics and includes classical electromagnetic zero-point radiation,
then one arrives at the Planck spectrum for classical radiation
equilibrium.\cite{B2018c} \ 

In work carried out more than forty years ago, it was shown\cite{B1976a} that
the addition of a nonlinear term to a harmonic oscillator led to a
nonrelativistic, nonlinear mechanical oscillator which scattered
electromagnetic radiation (treated in the radiation-dipole approximation)
toward the Rayleigh-Jeans spectrum. \ Thus the Rayleigh-Jeans spectrum was
stable under scattering by this nonrelativistic nonlinear system, but any
other spectrum of random classical radiation (including Lorentz-invariant
zero-point radiation) was changed by the scattering of this system, and the
radiation spectrum was pushed toward the Rayleigh-Jeans spectrum. \ The
crucial aspect is that the scattering system was a \textit{nonrelativistic}
classical system.\cite{vanVleck} \ 

In the calculation in the present article, we show that a simple harmonic
oscillator treated with its radiation multipole moments (not just the electric
dipole moment) leaves the spectrum of classical electromagnetic
\textit{zero-point} radiation invariant. \ Our calculation goes through only
the quadrupole order, but there are good reasons to expect the validity of the
scattering results to hold for all the multipole moments. \ Here the crucial
aspect is that harmonic oscillator motion in the small-amplitude regime is the
same in both relativistic and nonrelativistic physics, whereas the treatment
of electromagnetic multipole radiation emission and absorption is fully within
the relativistic regime. \ The spectrum of classical electromagnetic
zero-point radiation is invariant under Lorentz transformation, and we should
expect that the spectrum will be preserved only by a \textit{relativistic}
classical scattering system. \ Here we give the first example of a
\textit{relativistic scattering} calculation; indeed it involves equilibrium
at the \textit{zero-point} radiation spectrum. \ 

\subsubsection{Outline of the Article \ }

We start out by reviewing the treatment of random radiation within classical
physics. \ We follow this with a review of the exact steady-state behavior of
a point harmonic oscillator when located in a bath of random radiation. \ Then
we repeat the calculation in a form involving energy balance for energy
absorption and emission, since this is the form which will be used for the
quadrupole terms appearing for non-zero amplitude. \ After demonstrating that
the average oscillator amplitude can be obtained correctly from the energy
balance at the fundamental frequency of the oscillator, we turn to energy
balance at the first harmonic of the oscillator frequency when the oscillator
amplitude is non-zero. \ We first calculate the energy absorbed by the
oscillator from the random radiation spectrum during a short time interval
$\tau,$ and then we obtain the radiation emission. \ We find that energy
balance at the first harmonic requires that the radiation per normal mode at
the first harmonic should be double the energy per normal mode at the
fundamental. \ This is precisely the statement that the spectrum of random
radiation must correspond to the zero-point radiation spectrum which is linear
in frequency. \ Our calculation demonstrates explicitly that classical
scattering by this relativistic system does not lead to the Rayleigh-Jeans
spectrum but rather to Lorentz-invariant zero-point radiation. \ We then
explain why we expect the detailed balance to extend to all the radiation
harmonics of the harmonic-oscillator system. \ Finally, we comment upon our
current understanding of blackbody radiation within classical physics. \ 

\section{Set-Up for the Detailed-Balance Calculation}

\subsection{Random Classical Electromagnetic Radiation}

We consider a large volume $V$ containing source-free isotropic random
radiation. \ The random radiation present in the enclosure of volume $V$ can
be given as a sum over plane waves with periodic boundary
conditions\cite{B2016b}%
\begin{equation}
\mathbf{E}(\mathbf{r,}t)=\sum_{\mathbf{k}}\sum_{\lambda=1}^{2}%
\widehat{\epsilon}(\mathbf{k},\lambda)\mathfrak{h}(\omega)\cos[\mathbf{k}%
\cdot\mathbf{r}-\omega t+\theta(\mathbf{k},\lambda)] \label{EE}%
\end{equation}%
\begin{equation}
\mathbf{B}(\mathbf{r,}t)=\sum_{\mathbf{k}}\sum_{\lambda=1}^{2}\widehat{k}%
\times\widehat{\epsilon}(\mathbf{k},\lambda)\mathfrak{h}(\omega)\cos
[\mathbf{k}\cdot\mathbf{r}-\omega t+\theta(\mathbf{k},\lambda)] \label{BBB}%
\end{equation}
where the wave vector $\mathbf{k}$ takes the values $\mathbf{k=}%
\widehat{\mathbf{x}}l2\pi/a+\widehat{y}m2\pi/a+\widehat{z}n2\pi/a$ for
$l,~m,~n$ running over all positive and negative integers, the constant $a$ is
the length of a side of the box for periodic boundary conditions with volume
$V=a^{3},$ the random phases $\theta(\mathbf{k},\lambda)$ are distributed
independently and uniformly\cite{Rice} over the interval $(0,2\pi],$ and the
amplitude is given by
\begin{equation}
\mathfrak{h}(\omega)=\left(  \frac{8\pi\mathcal{E(}\omega\mathcal{)}}{a^{3}%
}\right)  ^{1/2} \label{h}%
\end{equation}
where $\mathcal{E}(\omega)$ is the energy per radiation normal mode of
frequency $\omega=ck=c|\mathbf{k}|.$ Because our calculations will seem
complicated, we will introduce compact notations.\cite{note} \ The random
electric and magnetic fields in Eqs. (\ref{EE}) and (\ref{BBB}) will be
written compactly as%
\begin{equation}
\mathbf{E}(\mathbf{r,}t)=\sum_{\mu,\mathbf{k,\lambda}}\widehat{\epsilon}%
\frac{\mathfrak{h}}{2}FRA \label{Ec}%
\end{equation}%
\begin{equation}
\mathbf{B}(\mathbf{r,}t)=\sum_{\mu,\mathbf{k,\lambda}}\widehat{k}%
\times\widehat{\epsilon}\frac{\mathfrak{h}}{2}FRA \label{Bc}%
\end{equation}
where in addition to the sums over $\mathbf{k}$ and $\lambda,$ we include the
sum over $\mu=\pm1$,%
\begin{equation}
\sum_{\mu}f(\mu)=f(-1)+f(+1)
\end{equation}
for any function $f(\mu).$ \ Here the abbreviated notation writes
$\widehat{\epsilon}$ for $\widehat{\epsilon}(\mathbf{k},\lambda),$
$\mathfrak{h}$ for $\mathfrak{h}(\omega),$ and
\begin{equation}
F=\exp[i\mu\mathbf{k\cdot r}], \label{F}%
\end{equation}%
\begin{equation}
R=\exp[-i\mu\omega t],
\end{equation}
and
\begin{equation}
A=\exp[i\mu\theta(\mathbf{k},\lambda)].
\end{equation}

\subsection{Dipole Oscillator in a Harmonic Oscillator Potential}

The charged particle of mass $m$ and charge $e$, constrained to move along the
$x$-axis in a harmonic potential, has the equation of motion (for small amplitude)%

\begin{equation}
m\ddot{x}=-m\omega_{0}^{2}x+m\Gamma\dddot{x}+eE_{x}(0,t) \label{dedip}%
\end{equation}
where the energy emission and absorption has been evaluated in the point
dipole radiation limit, and where $\Gamma=2e^{2}/(3mc^{3}).$ \ Here the
radiation damping term $m\Gamma\dddot{x}$ includes only the electric dipole
radiation emission, (and not the damping associated with the electric
quadrupole or higher multipole moments), and the driving radiation field
$\mathbf{E}(\mathbf{r},t)$ is evaluated at the center of the oscillator,
$\mathbf{r}=0$. \ This linear equation has been solved many times
before.\cite{B2016b} In our compact notation of Eq. (\ref{Ec}), the (real)
steady-state solution takes the form%
\begin{equation}
x(t)=\frac{e}{2m}\sum_{\mu,\mathbf{k,\lambda}}\epsilon_{x}\mathfrak{h}%
\frac{RA}{C} \label{xt}%
\end{equation}
where the symbol $C$ stands for
\begin{equation}
C=-\left(  \mu\omega\right)  ^{2}+\omega_{0}^{2}-i\Gamma\left(  \mu
\omega\right)  ^{3}. \label{CC}%
\end{equation}
The factor $F$ (appearing in Eq. (\ref{F})) does not appear in Eq. (\ref{xt})
because the driving electric field is evaluated at $\mathbf{r}=0.$ \ 

The average value of $x^{2}$ can be evaluated as
\begin{align}
\left\langle x^{2}\right\rangle  &  =\left\langle \left(  \frac{e}{2m}%
\sum_{\mu_{1},\mathbf{k}_{1}\mathbf{,\lambda}_{1}}\epsilon_{x1}\mathfrak{h}%
_{1}\frac{R_{1}A_{1}}{C_{1}}\right)  \left(  \frac{e}{2m}\sum_{\mu
_{2},\mathbf{k}_{2}\mathbf{,\lambda}_{2}}\epsilon_{x2}\mathfrak{h}_{2}%
\frac{R_{2}A_{2}}{C_{2}}\right)  \right\rangle \nonumber\\
&  =\left(  \frac{e}{2m}\right)  ^{2}\sum_{\mu_{1},\mathbf{k}_{1}%
\mathbf{,\lambda}_{1}}\sum_{\mu_{2},\mathbf{k}_{2}\mathbf{,\lambda}_{2}%
}\epsilon_{x1}\epsilon_{x2}\mathfrak{h}_{1}\mathfrak{h}_{2}\frac{R_{1}R_{2}%
}{C_{1}C_{2}}\left\langle A_{1}A_{2}\right\rangle \label{xtt}%
\end{align}
Now averaging over the random phases \ $\theta(\mathbf{k,}\lambda)$, we have
\begin{align}
\left\langle A_{1}A_{2}\right\rangle  &  =\left\langle \exp[i\mu_{1}%
\theta(\mathbf{k}_{1},\lambda_{1})]\exp[i\mu_{2}\theta(\mathbf{k}_{2}%
,\lambda_{2})]\right\rangle \nonumber\\
&  =\delta_{\mu_{1},-\mu_{2}}\delta_{\mathbf{k}_{1},\mathbf{k}_{2}}%
\delta_{\lambda_{1},\lambda_{2}}=\delta_{1(-2)} \label{AAav}%
\end{align}
where the last expression $\delta_{1(-2)}$ involves a short-hand notation.
\ Now summing over the Kronecker delta terms arising in Eq. (\ref{AAav}),
equation (\ref{xtt}) becomes%
\begin{align}
\left\langle x^{2}\right\rangle  &  =\left(  \frac{e}{2m}\right)  ^{2}%
\sum_{\mu_{1},\mathbf{k}_{1}\mathbf{,\lambda}_{1}}\epsilon_{x1}^{2}%
\mathfrak{h}_{1}^{2}\frac{R_{1}R_{(-1)}}{C_{1}C_{(-1)}}\nonumber\\
&  =2\left(  \frac{e}{2m}\right)  ^{2}\sum_{\mathbf{k}_{1}\mathbf{,\lambda
}_{1}}\epsilon_{x1}^{2}\mathfrak{h}_{1}^{2}\frac{1}{C_{1}C_{(-1)}} \label{xx}%
\end{align}
since $\widehat{\epsilon}(\mathbf{k},\lambda)$ and $\mathfrak{h}(\omega)$ are
independent of $\mu,$ while $R_{1}R_{(-1)}=1,$ and $C_{1}C_{(-1)}=(-\omega
^{2}-\omega_{0}^{2})^{2}+(\Gamma\omega^{3})^{2}.$ \ 

Next we assume that the distance $a$ used for the periodic boundary conditions
on the radiation is very large so that the sum can be approximated as an
integral,. \ Therefore we can write
\begin{equation}
k_{x}=l\frac{2\pi}{a},\text{ \ }k_{y}=m\frac{2\pi}{a},\text{ \ }k_{z}%
=n\frac{2\pi}{a},
\end{equation}
and
\begin{equation}
\sum_{\mathbf{\lambda}}\epsilon_{x}^{2}=\frac{k^{2}-k_{x}^{2}}{k^{2}},
\end{equation}
and so approximate the sums as%
\begin{equation}
\sum_{\mathbf{k,\lambda}}\epsilon_{x}^{2}=\sum_{l,m,n}\sum_{\mathbf{\lambda}%
}\epsilon_{x}^{2}\approx\left(  \frac{a}{2\pi}\right)  ^{3}\int d^{3}%
k\sum_{\mathbf{\lambda}}\epsilon_{x}^{2}=\left(  \frac{a}{2\pi}\right)
^{3}\int d^{3}k\frac{(k^{2}-k_{x}^{2})}{k^{2}}. \label{SI0}%
\end{equation}
Choosing $x$ as the polar axis, the needed angular integration over
$\mathbf{k}$ involves
\begin{equation}
\int d\Omega(1-\cos^{2}\theta)=\int_{0}^{2\pi}d\phi\int_{0}^{\pi}d\theta
\sin\theta(1-\cos^{2}\theta)=8\pi/3. \label{ang}%
\end{equation}
Then using Eqs. (\ref{h}), (\ref{CC}), (\ref{SI0}),and (\ref{ang}), we have
\begin{align}
\left\langle x^{2}\right\rangle  &  =2\left(  \frac{e}{2m}\right)  ^{2}\left(
\frac{a}{2\pi}\right)  ^{3}\int_{0}^{\infty}\frac{d\omega\omega^{2}}{c^{3}%
}\frac{8\pi}{3}\left(  \frac{8\pi\mathcal{E(}\omega\mathcal{)}}{a^{3}}\right)
\frac{1}{(-\omega^{2}+\omega_{0}^{2})^{2}+(\Gamma\omega^{3})^{2}}\nonumber\\
&  \approx\frac{4}{3\pi}\left(  \frac{e}{m}\right)  ^{2}\frac{\omega_{0}%
^{2}\mathcal{E(}\omega_{0}\mathcal{)}}{c^{3}}\int_{0}^{\infty}d\omega\frac
{1}{(2\omega_{0})^{2}(-\omega+\omega_{0})^{2}+(\Gamma\omega^{3})^{2}%
}\nonumber\\
&  =\frac{4}{3\pi}\left(  \frac{e}{m}\right)  ^{2}\frac{\omega_{0}%
^{2}\mathcal{E(}\omega_{0}\mathcal{)}}{c^{3}}\frac{\pi}{2\omega_{0}^{4}%
}\left(  \frac{3mc^{3}}{2e^{2}}\right)  =\frac{\mathcal{E(}\omega
_{0}\mathcal{)}}{m\omega_{0}^{2}}\nonumber
\end{align}
where we have assumed that the integrand is sharply peaked at $\omega_{0}$ and
therefore have substituted $\omega_{0}$ for every appearance of $\omega,$
except where the combination $\omega-\omega_{0}$ occurs. \ Also, we have
evaluated the integral as%
\begin{align}
\int_{0}^{\infty}d\omega\frac{1}{(2\omega_{0})^{2}(-\omega+\omega_{0}%
)^{2}+(\Gamma\omega^{3})^{2}}  &  \approx\int_{-\infty}^{\infty}d\omega
\frac{1}{(2\omega_{0})^{2}(-\omega+\omega_{0})^{2}+(\Gamma\omega_{0}^{3})^{2}%
}\nonumber\\
&  =\frac{\pi}{2\omega_{0}\Gamma\omega_{0}^{3}}=\frac{\pi}{2\omega_{0}^{4}%
}\left(  \frac{3mc^{3}}{2e^{2}}\right)
\end{align}
using%
\begin{equation}
\int_{-\infty}^{\infty}\frac{dx}{a^{2}x^{2}+b^{2}}=\frac{\pi}{ab}. \label{INT}%
\end{equation}
Thus we find%
\begin{equation}
\left\langle x^{2}\right\rangle =\frac{\mathcal{E(}\omega_{0}\mathcal{)}%
}{m\omega_{0}^{2}}. \label{x2av}%
\end{equation}
We notice that the average amplitude of the motion decreases as the mass $m$
increases for fixed $\omega_{0}.$ \ 

\section{General Scattering Calculation}

The steady-state solution in Eq. (\ref{xt}) is exact for the linear equation
given in Eq. (\ref{dedip}). \ However, equation (\ref{dedip}) is only an
approximation to the actual physical situation because it ignores the non-zero
amplitude of the oscillator motion. \ Often one speaks of a \textquotedblleft
point\textquotedblright\ dipole oscillator. \ The condition for the
\textquotedblleft relativistic\textquotedblright\ compatibility of the motion
is that $|\omega_{0}x|<<c$ so that the charged particle speed is small
compared to the speed of light. \ However, even if this conditon is satisfied,
equation (\ref{dedip}) still ignores the radiation emission and absorption
associated with the non-zero excursion of the charged particle. The solution
in Eq. (\ref{xt})\ holds for a point dipole oscillator whose excursion is so
small that the dipole approximations may be applied consistently and radiation
at all the higher harmonics can be ignored. \ We wish to show that this same
expression given in Eq. (\ref{xt}) holds even when we consider the radiation
emission and absorption associated with a finite non-zero excursion of the
oscillator where the radiation at the harmonics is treated in detail. \ 

In order to proceed with the more complicated analysis, we first simplify.
\ We drop the radiation damping term (to be treated later) in Eq.
(\ref{dedip}), and consider only the driving by the random radiation where we
now include the first correction in our approximation, so that%
\begin{equation}
E_{x}(\widehat{i}x,t)\approx E_{x}(0,t)+x[\partial_{x^{\prime}}E_{x}%
(\widehat{i}x^{\prime},t)]_{x^{\prime}=0}+...,
\end{equation}
giving an equation of motion for the oscillator%
\begin{equation}
\ddot{x}=-\omega_{0}^{2}x+(e/m)\left\{  E_{x}(0,t)+x[\partial_{x^{\prime}%
}E_{x}(\widehat{i}x^{\prime},t)]_{x^{\prime}=0}+...\right\}  .\label{DE2}%
\end{equation}
We will use this equation to calculate the energy absorbed by the oscillator
from the radiation during a short time interval $\tau$, and then will treat
separately the energy emitted by the oscillator into the radiation field. \ 

We write a series expansion for the oscillator displacement
\begin{equation}
x=x_{1}+x_{2}+x_{3}+...
\end{equation}
where the subscript refers to the number of factors of $\mathfrak{h}$ which
appear in the steady-state expression. \ Substituting this expansion into Eq.
(\ref{DE2}), we can separate the terms depending on the number of powers of
$\mathfrak{h}$. \ In first order in $\mathfrak{h}$, we have%
\begin{equation}
\ddot{x}_{1}=-\omega_{0}^{2}x_{1}+(e/m)E_{x}(0,t), \label{de1}%
\end{equation}
whereas in second order, we find
\begin{equation}
\ddot{x}_{2}=-\omega_{0}^{2}x_{2}+(e/m)x[\partial_{x^{\prime}}E_{x}%
(\widehat{i}x^{\prime},t)]_{x^{\prime}=0}. \label{de2}%
\end{equation}

\subsection{Energy Balance Calculation for the Dipole Terms}

Equation (\ref{de1}) corresponds to the same approximations as in Eq.
(\ref{dedip}) except that the radiation damping term has been omitted. \ Thus
equation (\ref{de1}) has no energy-loss mechanism while it is driven by the
forcing term. \ The solution of the differential equation (\ref{de1}) will
involve the sum $x_{1}=x_{1c}+x_{1p}$ of a particular solution $x_{1p}$ of the
full equation (\ref{de1}) and a complementary solution $x_{1c}$ of the
homogeneous equation, chosen so as to meet the initial conditions for the
displacement and velocity of the oscillator at time $t=0.$ \ We are interested
in energy balance, and so we will calculate the average amount of energy
delivered to the oscillator by the forcing electromagnetic field during a
short time interval $\tau,$ and will then balance this energy against the
energy radiated by the oscillator during this same short time interval $\tau.$
\ Both the amount of energy absorbed and the amount of energy emitted by the
oscillator may depend upon the displacement and velocity of the oscillator at
time $t=0.$ \ Since we have the solution for the oscillator motion in the
dipole approximation, we will take the initial conditions from the full
solution given in Eq. (\ref{dedip}). \ Of course, since we have the full
solution for the dipole approximation, we could use this solution directly
when calculating the energy absorption and emission in the dipole case.
\ However, since in calculating the quadrupole energy absorption and emission,
we will not have the exact solution, it seems wise to show the consistency of
the basic procedure even though this is redundant in the dipole case. \ 

Provided $\omega\neq\omega_{0,}$ a particular solution of Eq. (\ref{de1})
corresponds to the steady-state solution obtained from the driving field in
Eq. (\ref{EE}),%
\begin{equation}
x_{1p}=\frac{e}{2m}\sum_{\mu,\mathbf{k,\lambda}}\epsilon_{x}\mathfrak{h}%
\frac{RA}{D} \label{x1p}%
\end{equation}
where
\begin{equation}
D=-(\mu\omega)^{2}+\omega_{0}^{2}.
\end{equation}
The summand in Eq. (\ref{x1p}) diverges at $\omega=\omega_{0}.$ The
complementary solution to Eq. (\ref{de1}) is a solution to the homogeneous
equation, so that
\begin{equation}
x_{1c}=X\cos\omega_{0}t+Y\sin\omega_{0}t=\frac{1}{2}\sum_{\mu_{0}}\left(
X+i\mu_{0}Y\right)  R_{0} \label{x1c}%
\end{equation}
where $R_{0}=\exp[-i\mu_{0}\omega_{0}t].$ \ The constants $X$ and $Y$ are
determined such that at time $t=0,$ there is agreement with Eq. (\ref{xt}), so%

\begin{equation}
x_{1c}(0)=x(0)-x_{1p}(0)
\end{equation}
and
\begin{equation}
\dot{x}_{1c}(0)=\dot{x}(0)-\dot{x}_{1p}(0),
\end{equation}
However, the terms $x(0)$ and $\dot{x}(0)$ from Eq. (\ref{xt}) are finite at
$\omega=\omega_{0}$ and will not contribute to resonant energy pick-up from
the driving electric field. \ Therefore these terms can be omitted. \ Thus we
need only $x_{1c}(0)\approx-x_{1p}(0)$ and $\dot{x}_{1c}(0)\approx-\dot
{x}_{1p}(0)$ from Eqs. (\ref{x1p}) and (\ref{x1c}), giving
\begin{equation}
X=-\frac{e}{2m}\sum_{\mu,\mathbf{k,\lambda}}\epsilon_{x}\mathfrak{h}\frac
{A}{D}%
\end{equation}
and%
\begin{equation}
\omega_{0}Y=-\frac{e}{2m}\sum_{\mu,\mathbf{k,\lambda}}\epsilon_{x}%
\mathfrak{h(-i\mu}\omega)\frac{A}{D}.
\end{equation}
Then the complementary solution in Eq. (\ref{x1c}) can be written as
\begin{align}
x_{c}  &  =\frac{1}{2}\sum_{\mu_{0}}\left[  \left(  -\frac{e}{2m}\sum
_{\mu,\mathbf{k,\lambda}}\epsilon_{x}\mathfrak{h}\frac{A}{D}\right)  +i\mu
_{0}\left(  -\frac{e}{2m}\sum_{\mu,\mathbf{k,\lambda}}\epsilon_{x}%
\mathfrak{h}\frac{\mathfrak{(-i\mu}\omega)}{\omega_{0}}\frac{A}{D}\right)
\right]  R_{0}\nonumber\\
&  =\frac{-1}{2}\frac{e}{2m}\sum_{\mu_{0}}\sum_{\mu,\mathbf{k,\lambda}%
}\epsilon_{x}\mathfrak{h}\frac{A}{D}\left(  1+\frac{\mu_{0}\mu\omega}%
{\omega_{0}}\right)  R_{0}. \label{xcc}%
\end{align}

The energy absorbed from the random radiation in the $x_{1}$ approximation for
the oscillator displacement is
\begin{align}
W_{1}  &  =\int_{0}^{\tau}dt\dot{x}_{1}eE_{x}(0,t)=\int_{0}^{\tau}dt(\dot
{x}_{1c}+\dot{x}_{1p})eE_{x}(0,t)\nonumber\\
&  =\int_{0}^{\tau}dt\dot{x}_{1c}eE_{x}(0,t)+\int_{0}^{\tau}dt\dot{x}%
_{1p}eE_{x}(0,t). \label{Wdi1}%
\end{align}
We first consider the last term involving $\dot{x}_{1p}$ in Eq. (\ref{x1p})
and $E_{x}(0,t)~$in Eq. (\ref{Ec}). $\ $Taking the average over the random
phases, we are dealing with%
\begin{align}
\left\langle \int_{0}^{\tau}dt\dot{x}_{1p}eE_{x}(0,t)\right\rangle  &
=\left\langle \int_{0}^{\tau}dt\left(  \frac{e}{2m}\sum_{\mu_{1}%
,\mathbf{k}_{1}\mathbf{,\lambda}_{1}}\epsilon_{x1}\mathfrak{h}_{1}(-i\mu
_{1}\omega_{1})\frac{R_{1}A_{1}}{D_{1}}\right)  e\left(  \sum_{\mu
_{1},\mathbf{k}_{2}\mathbf{,\lambda}_{2}}\widehat{\epsilon}_{2}\frac
{\mathfrak{h}_{2}}{2}R_{2}A_{2}\right)  \right\rangle \nonumber\\
&  =\int_{0}^{\tau}dt\frac{e^{2}}{2m}\sum_{\mu_{1},\mathbf{k}_{1}%
\mathbf{,\lambda}_{1}}\sum_{\mu_{1},\mathbf{k}_{2}\mathbf{,\lambda}_{2}%
}\epsilon_{x1}\epsilon_{x2}\mathfrak{h}_{1}\frac{\mathfrak{h}_{2}}{2}%
(-i\mu_{1}\omega_{1})\frac{R_{1}R_{2}}{D_{1}}\left\langle A_{1}A_{2}%
\right\rangle
\end{align}
where $\left\langle A_{1}A_{2}\right\rangle $ is given in Eq.\ (\ref{AAav}).
\ Now summing over the second set of indices, we have%
\begin{equation}
\left\langle \int_{0}^{\tau}dt\dot{x}_{1p}eE_{x}(0,t)\right\rangle =\int%
_{0}^{\tau}dt\frac{e^{2}}{2m}\sum_{\mu_{1},\mathbf{k}_{1}\mathbf{,\lambda}%
_{1}}\epsilon_{x1}^{2}\mathfrak{h}_{1}^{2}\frac{(-i\mu_{1}\omega_{1})}{2D_{1}%
}=0, \label{Zero}%
\end{equation}
since $\epsilon_{x1}^{2}$ and $\mathfrak{h}_{1}^{2}$are independent of
$\mu_{1}$ while the expression in Eq. (\ref{Zero}) contains exactly one factor
of $\mu_{1}=\pm1.$ \ 

The average over the random phase for the term involving $\dot{x}_{c}$ in Eq.
(\ref{Wdi1}) involves Eqs. (\ref{xcc}) and (\ref{Ec}), giving
\begin{align}
\left\langle W_{1}\right\rangle  &  =\left\langle \int_{0}^{\tau}dt\dot
{x}_{1c}eE_{x}(0,t)\right\rangle \nonumber\\
&  =\left\langle \int_{0}^{\tau}dt\left(  \frac{-1}{2}\frac{e}{2m}\sum
_{\mu_{0}}\sum_{\mu_{1},\mathbf{k}_{1}\mathbf{,\lambda}_{1}}\epsilon
_{x1}\mathfrak{h}_{1}\frac{A_{1}}{D_{1}}\left(  1+\frac{\mu_{0}\mu_{1}%
\omega_{1}}{\omega_{0}}\right)  (-i\mu_{0}\omega_{0})R_{0}\right)  \right.
\nonumber\\
&  \left.  \times e\left(  \sum_{\mu_{1},\mathbf{k}_{2}\mathbf{,\lambda}_{2}%
}\widehat{\epsilon}_{2}\frac{\mathfrak{h}_{2}}{2}R_{2}A_{2}\right)
\right\rangle \\
&  =\frac{-1}{2}\int_{0}^{\tau}dt\frac{e^{2}}{2m}\sum_{\mu_{0}}\sum_{\mu
_{1},\mathbf{k}_{1}\mathbf{,\lambda}_{1}}\frac{\epsilon_{x1}\mathfrak{h}_{1}%
}{D_{1}}\left(  1+\frac{\mu_{0}\mu_{1}\omega_{1}}{\omega_{0}}\right)
(-i\mu_{0}\omega_{0})\sum_{\mu_{1},\mathbf{k}_{2}\mathbf{,\lambda}_{2}%
}\widehat{\epsilon}_{2}\frac{\mathfrak{h}_{2}}{2}R_{0}R_{2}\left\langle
A_{1}A_{2}\right\rangle \nonumber\\
&  =\frac{-1}{4}\frac{e^{2}}{2m}\sum_{\mu_{0}}\sum_{\mu_{1},\mathbf{k}%
_{1}\mathbf{,\lambda}_{1}}\frac{\epsilon_{x1}^{2}\mathfrak{h}_{1}^{2}}{D_{1}%
}\left(  1+\frac{\mu_{0}\mu_{1}\omega_{1}}{\omega_{0}}\right)  (-i\mu
_{0}\omega_{0})\int_{0}^{\tau}dtR_{0}R_{(-1)} \label{W1avt}%
\end{align}
where we have averaged over the random phases as in Eq. (\ref{AAav}) and then
summed over $\mu_{2},$~$\mathbf{k}_{2},$ and $\lambda_{2}.$ \ Now we need the
integral%
\begin{align}
\int_{0}^{\tau}dtR_{0}R_{(-1)}  &  =\int_{0}^{\tau}dt\exp[-i(\mu_{0}\omega
_{0}-\mu_{1}\omega_{1})t]\nonumber\\
&  =\frac{1-\exp[-i(\mu_{0}\omega_{0}-\mu_{1}\omega_{1})\tau]}{i(\mu_{0}%
\omega_{0}-\mu_{1}\omega_{1})}. \label{RRint}%
\end{align}
Then combining Eqs. (\ref{W1avt}) and (\ref{RRint}), we have%
\begin{equation}
\left\langle W_{1}\right\rangle =\frac{1}{4}\frac{e^{2}}{2m}\sum
_{\mathbf{k}_{1}\mathbf{,\lambda}_{1}}\frac{\epsilon_{x1}^{2}\mathfrak{h}%
_{1}^{2}}{D_{1}}\sum_{\mu_{0}}\sum_{\mu_{1}}\left(  \omega_{0}+\mu_{0}\mu
_{1}\omega_{1}\right)  \mu_{0}\frac{1-\exp[-i(\mu_{0}\omega_{0}-\mu_{1}%
\omega_{1})\tau]}{(\mu_{0}\omega_{0}-\mu_{1}\omega_{1})}.
\end{equation}
Only if $\mu_{0}=\mu_{1}$ will we have a resonant denominator involving
$\omega-\omega_{0}$. \ Therefore retaining only the resonant terms, we write%
\begin{align}
\left\langle W_{1}\right\rangle  &  =\frac{1}{4}\frac{e^{2}}{2m}%
\sum_{\mathbf{k}_{1}\mathbf{,\lambda}_{1}}\frac{\epsilon_{x1}^{2}%
\mathfrak{h}_{1}^{2}}{D_{1}}\sum_{\mu_{0}}\left(  \omega_{0}+\omega
_{1}\right)  \frac{1-\exp[-i(\mu_{0}\omega_{0}-\mu_{0}\omega_{1})\tau
]}{(\omega_{0}-\omega_{1})}\\
&  =\frac{1}{2}\frac{e^{2}}{2m}\sum_{\mathbf{k}_{1}\mathbf{,\lambda}_{1}%
}\epsilon_{x1}^{2}\mathfrak{h}_{1}^{2}\frac{1-\cos[(\omega_{0}-\omega_{1}%
)\tau]}{(\omega_{0}-\omega_{1})^{2}},
\end{align}
\ where we have used%
\begin{equation}
\sum_{\mu_{0}}\frac{1-\exp[-i(\mu_{0}\omega_{0}-\mu_{0}\omega_{1})\tau
]}{(\omega_{0}-\omega_{1})}=\frac{2-2\cos[(\omega_{0}-\omega_{1})\tau
]}{(\omega_{0}-\omega_{1})}.
\end{equation}
Now we need to use Eqs. (\ref{SI0}) and (\ref{ang}) so as to convert from sums
to integrals. \ Then we have%
\begin{align}
\left\langle W_{1}\right\rangle  &  =\frac{1}{2}\frac{e^{2}}{2m}\left(
\frac{a}{2\pi}\right)  ^{3}\int_{0}^{\infty}dkk^{2}\frac{8\pi}{3}\left(
\frac{8\pi\mathcal{E(}\omega\mathcal{)}}{a^{3}}\right)  \frac{1-\cos
[(\omega_{0}-\omega_{1})\tau]}{(\omega_{0}-\omega_{1})^{2}}\nonumber\\
&  \approx\frac{1}{2}\frac{e^{2}}{2m}\left(  \frac{a}{2\pi}\right)  ^{3}%
\frac{1}{c^{3}}\int_{-\infty}^{\infty}d\omega\omega_{0}^{2}\frac{8\pi}%
{3}\left(  \frac{8\pi\mathcal{E(}\omega_{0}\mathcal{)}}{a^{3}}\right)
\frac{1-\cos[(\omega_{0}-\omega_{1})\tau]}{(\omega_{0}-\omega_{1})^{2}%
}\nonumber\\
&  =\frac{1}{2}\frac{e^{2}}{2m}\left(  \frac{a}{2\pi}\right)  ^{3}\frac
{1}{c^{3}}\omega_{0}^{2}\frac{8\pi}{3}\left(  \frac{8\pi\mathcal{E(}\omega
_{0}\mathcal{)}}{a^{3}}\right)  \pi\tau\nonumber\\
&  =\frac{2}{3}\frac{e^{2}}{m}\frac{1}{c^{3}}\omega_{0}^{2}\left(
\frac{\mathcal{E(}\omega_{0}\mathcal{)}}{1}\right)  \tau\label{W1}%
\end{align}
where we have used\cite{GR432}%
\begin{equation}
\int_{-\infty}^{\infty}dx\frac{1-\cos(x\tau)}{x^{2}}=\pi\tau. \label{GR432}%
\end{equation}

The average energy $\left\langle W_{\mathbf{p}}\right\rangle $ radiated by an
electric dipole during a time $\tau$ is given by Larmor's formula%
\begin{equation}
\left\langle W_{\mathbf{p}}\right\rangle =\frac{2}{3}\frac{e^{2}}{c^{3}}%
\omega_{0}^{4}\left\langle x^{2}\right\rangle \tau. \label{Wpp}%
\end{equation}
Comparing the energy pick-up in Eq. (\ref{W1}) and loss in Eq. (\ref{Wpp}), we
have
\begin{equation}
\frac{2}{3}\frac{e^{2}}{m}\frac{1}{c^{3}}\omega_{0}^{2}\left(  \frac
{\mathcal{E(}\omega_{0}\mathcal{)}}{1}\right)  \tau=\frac{2}{3}\frac{e^{2}%
}{c^{3}}\omega_{0}^{4}\left\langle x^{2}\right\rangle \tau. \label{Bal1}%
\end{equation}
or%
\begin{equation}
\left\langle x^{2}\right\rangle =\frac{\mathcal{E(}\omega_{0}\mathcal{)}%
}{m\omega_{0}^{2}},
\end{equation}
exactly as found above in Eq. (\ref{x2av}). \ 

\subsection{Energy Balance for the Quadrupole Terms}

Now we treat the energy balance involving the quadrupole moment when the
oscillator behavior is given by the same random motion as was determined using
the dipole approximation for radiation. \ We now consider the terms in Eq.
(\ref{DE2}) which involve two factors of $\mathfrak{h,}$ arising from Eq.
(\ref{xt}) and Eq. (\ref{EE}),%

\begin{align}
\ddot{x}_{2}  &  =-\omega_{0}^{2}x_{2}+(e/m)x[\partial_{x}E_{x}(\widehat{i}%
x,t)]_{x=0}.\nonumber\\
&  =-\omega_{0}^{2}x_{2}+\frac{e}{m}\left(  \frac{e}{2m}\sum_{\mu
_{1},\mathbf{k}_{1}\mathbf{,\lambda}_{1}}\epsilon_{x1}\mathfrak{h}_{1}%
\frac{R_{1}A_{1}}{C_{1}}\right)  \left[  \sum_{\mu_{2},\mathbf{k}%
_{2}\mathbf{,\lambda}_{2}}\epsilon_{x2}\frac{\mathfrak{h}_{2}}{2}(i\mu
_{2}k_{x2})R_{2}A_{2}\right]
\end{align}
Once again, we can solve this differential equation in terms of the sum of a
particular solution $x_{2p}$ and a complementary solution $x_{2c}$ of the
homogeneous equation. \ \ From the time dependence in $R_{1}$ and $R_{2},$ we
can obtain the particular solution $x_{2p}$ as%
\begin{equation}
x_{2p}=\left(  \frac{e}{2m}\right)  ^{2}\sum_{\mu_{1},\mathbf{k}%
_{1}\mathbf{,\lambda}_{1}}\sum_{\mu_{2},\mathbf{k}_{2}\mathbf{,\lambda}_{2}%
}\epsilon_{x1}\mathfrak{h}_{1}\epsilon_{x2}\mathfrak{h}_{2}(i\mu_{2}%
k_{x2})\frac{R_{1}R_{2}}{C_{1}D_{1+2}}A_{1}A_{2} \label{x2pp}%
\end{equation}
where%
\begin{equation}
D_{1+2}=-(\mu_{1}\omega_{1}+\mu_{2}\omega_{2})^{2}+\omega_{0}^{2}.
\end{equation}
The complementary solution involves $x_{2c}(t)=\mathcal{X}\cos\omega
_{0}t+\mathcal{Y}\sin\omega_{0}t,$ when $x_{2}(0)=x_{2c}(0)+x_{2p}(0)$ and
$\dot{x}_{2}(0)=\dot{x}_{2c}(0)+\dot{x}_{2p}(0).$ \ Now we are taking the
oscillator behavior as unchanged from the radiation dipole behavior so that
$x_{2}(0)=0$ and $\dot{x}_{2}(0)=0,$ so therefore $x_{2c}(0)=-x_{2p}$ and
$\dot{x}_{2c}(0)=-\dot{x}_{2p}.$ \ Then noting%
\begin{equation}
x_{2c}=\mathcal{X}\cos\omega_{0}t+\mathcal{Y}\sin\omega_{0}t=\frac{1}{2}%
\sum_{\mu_{0}}\left(  \mathcal{X}+i\mu_{0}\mathcal{Y}\right)  R_{0},
\end{equation}
we find%
\begin{equation}
\mathcal{X}=-\left(  \frac{e}{2m}\right)  ^{2}\sum_{\mu_{1},\mathbf{k}%
_{1}\mathbf{,\lambda}_{1}}\sum_{\mu_{2},\mathbf{k}_{2}\mathbf{,\lambda}_{2}%
}\epsilon_{x1}\mathfrak{h}_{1}\epsilon_{x2}\mathfrak{h}_{2}(i\mu_{2}%
k_{x2})\frac{A_{1}A_{2}}{C_{1}D_{1+2}}%
\end{equation}
and%
\begin{equation}
\omega_{0}\mathcal{Y}=-\left(  \frac{e}{2m}\right)  ^{2}\sum_{\mu
_{1},\mathbf{k}_{1}\mathbf{,\lambda}_{1}}\sum_{\mu_{2},\mathbf{k}%
_{2}\mathbf{,\lambda}_{2}}\epsilon_{x1}\mathfrak{h}_{1}\epsilon_{x2}%
\mathfrak{h}_{2}(i\mu_{2}k_{x2})[-i(\mu_{1}\omega_{1}+\mu_{2}\omega_{2}%
)]\frac{A_{1}A_{2}}{C_{1}D_{1+2}},
\end{equation}
so that%

\begin{equation}
x_{2c}=-\frac{1}{2}\sum_{\mu_{0}}\left(  \frac{e}{2m}\right)  ^{2}\sum
_{\mu_{1},\mathbf{k}_{1}\mathbf{,\lambda}_{1}}\sum_{\mu_{2},\mathbf{k}%
_{2}\mathbf{,\lambda}_{2}}\epsilon_{x1}\mathfrak{h}_{1}\epsilon_{x2}%
\mathfrak{h}_{2}(i\mu_{2}k_{x2})\left[  1+\mu_{0}\frac{(\mu_{1}\omega_{1}%
+\mu_{2}\omega_{2})}{\omega_{0}}\right]  \frac{A_{1}A_{2}}{C_{1}D_{1+2}}R_{0}%
\end{equation}
and
\begin{align}
\dot{x}_{2c}  &  =-\frac{1}{2}\sum_{\mu_{0}}\left(  \frac{e}{2m}\right)
^{2}\sum_{\mu_{1},\mathbf{k}_{1}\mathbf{,\lambda}_{1}}\sum_{\mu_{2}%
,\mathbf{k}_{2}\mathbf{,\lambda}_{2}}\epsilon_{x1}\mathfrak{h}_{1}%
\epsilon_{x2}\mathfrak{h}_{2}(i\mu_{2}k_{x2})\left[  1+\mu_{0}\frac{(\mu
_{1}\omega_{1}+\mu_{2}\omega_{2})}{\omega_{0}}\right] \nonumber\\
&  \times\frac{A_{1}A_{2}}{C_{1}D_{1+2}}(-i\mu_{0}\omega_{0})R_{0}.
\label{x2cc}%
\end{align}

We want to find the average power absorbed by the oscillator in connection
with $x_{2}$ during the short time interval $\tau,$
\begin{align}
\left\langle W_{2}\right\rangle  &  =\left\langle \int_{0}^{\tau}dt\dot{x}%
_{2}eE_{x}(\widehat{i}x,t)\right\rangle \nonumber\\
&  =\left\langle \int_{0}^{\tau}dt\dot{x}_{2}e[E_{x}(0,t)+x[\partial
_{x^{\prime}}E_{x}(\widehat{i}x^{\prime},t)]_{x^{\prime}=0}+...,]\right\rangle
\nonumber\\
&  =\left\langle \int_{0}^{\tau}dt\dot{x}_{2}ex[\partial_{x^{\prime}}%
E_{x}(\widehat{i}x^{\prime},t)]_{x^{\prime}=0}\right\rangle \nonumber\\
&  =\left\langle \int_{0}^{\tau}dt(\dot{x}_{2c}+\dot{x}_{2p})ex[\partial
_{x^{\prime}}E_{x}(\widehat{i}x^{\prime},t)]_{x^{\prime}=0}\right\rangle
\label{W2all}%
\end{align}
where we have dropped the term $\left\langle \int_{0}^{\tau}dt\dot{x}%
_{2}eE_{x}(0,t)\right\rangle $ because it involves an odd number of random
phases $A$ and hence vanishes. \ Now the term $\left\langle \int_{0}^{\tau
}dt(\dot{x}_{2p})ex[\partial_{x}E_{x}(\widehat{i}x,t)]_{x=0}\right\rangle $
\ in Eq.(\ref{W2all}) also vanishes because it involves four sums over $\mu$
but an odd number of factors of $\mu$: one factor of $\mu$ appearing in Eq.
(\ref{x2pp}), one factor of $\mu$ appearing from the time derivative, and a
third factor of $\mu$ appearing from the spatial derivative of the electric
field. \ The calculation is analogous to that for the term $\left\langle
\int_{0}^{\tau}dt\dot{x}_{1p}eE_{x}(0,t)\right\rangle $ which vanished in Eq.
(\ref{Zero}). \ 

This leaves
\begin{equation}
\left\langle W_{2}\right\rangle =\left\langle \int_{0}^{\tau}dt(\dot{x}%
_{2c})ex[\partial_{x^{\prime}}E_{x}(\widehat{i}x^{\prime},t)]_{x^{\prime}%
=0}\right\rangle \label{W2gen}%
\end{equation}
where%
\begin{align}
&  ex[\partial_{x^{\prime}}E_{x}(\widehat{x}x^{\prime},t)]_{x^{\prime}%
=0}\nonumber\\
&  =e\left(  \frac{e}{2m}\sum_{\mu_{3},\mathbf{k}_{3}\mathbf{,\lambda}_{4}%
}\epsilon_{x3}\mathfrak{h}_{3}\frac{R_{3}A_{3}}{C_{3}}\right)  \left[
\sum_{\mu_{4},\mathbf{k}_{4}\mathbf{,\lambda}_{4}}\epsilon_{x4}\frac
{\mathfrak{h}_{4}}{2}(i\mu_{4}k_{x4})R_{4}A_{4}\right]  . \label{XE2}%
\end{align}

Introducing Eqs. (\ref{x2cc}) and (\ref{XE2}) into Eq. (\ref{W2gen}), we
obtain an expression with an odd number of sums over $\mu$ and an odd number
of factors of $\mu,$%
\begin{align}
\left\langle W_{2}\right\rangle  &  =\left\langle \int_{0}^{\tau}dt\left(
-\frac{1}{2}\right)  \sum_{\mu_{0}}\left(  \frac{e}{2m}\right)  ^{2}\sum
_{\mu_{1},\mathbf{k}_{1}\mathbf{,\lambda}_{1}}\sum_{\mu_{2},\mathbf{k}%
_{2}\mathbf{,\lambda}_{2}}\epsilon_{x1}\mathfrak{h}_{1}\epsilon_{x2}%
\mathfrak{h}_{2}(i\mu_{2}k_{x2})\left[  1+\mu_{0}\frac{(\mu_{1}\omega_{1}%
+\mu_{2}\omega_{2})}{\omega_{0}}\right]  \right. \nonumber\\
&  \times\frac{A_{1}A_{2}}{C_{1}D_{1+2}}(-i\mu_{0}\omega_{0})R_{0}\left.
e\left(  \frac{e}{2m}\sum_{\mu_{3},\mathbf{k}_{3}\mathbf{,\lambda}_{4}%
}\epsilon_{x3}\mathfrak{h}_{3}\frac{R_{3}A_{3}}{C_{3}}\right)  \left[
\sum_{\mu_{4},\mathbf{k}_{4}\mathbf{,\lambda}_{4}}\epsilon_{x4}\frac
{\mathfrak{h}_{4}}{2}(i\mu_{4}k_{x4})R_{4}A_{4}\right]  \right\rangle
\nonumber\\
&  =\left(  -\frac{1}{4}\right)  \sum_{\mu_{0}}e\left(  \frac{e}{2m}\right)
^{3}\sum_{\mu_{1},\mathbf{k}_{1}\mathbf{,\lambda}_{1}}\sum_{\mu_{2}%
,\mathbf{k}_{2}\mathbf{,\lambda}_{2}}\sum_{\mu_{3},\mathbf{k}_{3}%
\mathbf{,\lambda}_{4}}\sum_{\mu_{4},\mathbf{k}_{4}\mathbf{,\lambda}_{4}%
}\epsilon_{x1}\epsilon_{x2}\epsilon_{x3}\epsilon_{x4}\mathfrak{h}%
_{1}\mathfrak{h}_{2}\mathfrak{h}_{3}\mathfrak{h}_{4}(i\mu_{2}k_{x2})(i\mu
_{4}k_{x4})\nonumber\\
&  \times\left[  1+\mu_{0}\frac{(\mu_{1}\omega_{1}+\mu_{2}\omega_{2})}%
{\omega_{0}}\right]  \frac{(-i\mu_{0}\omega_{0})}{C_{1}C_{3}D_{1+2}%
}\left\langle A_{1}A_{2}A_{3}A_{4}\right\rangle \int_{0}^{\tau}dtR_{0}%
R_{3}R_{4}%
\end{align}

Now averaging over the random phases, we have%
\begin{equation}
\left\langle A_{1}A_{2}A_{3}A_{4}\right\rangle =\delta_{1(-2)}\delta
_{3(-4)}+\delta_{1(-3)}\delta_{2(-4)}+\delta_{1(-4)}\delta_{2(-3)}.
\label{4As}%
\end{equation}
Resonant behavior will occur only when the random phase $\theta_{1}$ is
matched with $\theta_{3}$ and the phase $\theta_{2}$ is matched with
$\theta_{4},$ so that the only term which is resonant and hence of interest is
the center term $\delta_{1(-3)}\delta_{2(-4)}$ in Eq. (\ref{4As}). \ Taking
only this center term and summing over the indices labeled $3$ and $4$, while
noting \ that $\epsilon_{x}$ and $\mathfrak{h}$ do not depend on $\mu,$ while
$\mu_{2}\mu_{(-2)}=-1,$ we have%
\begin{align}
\left\langle W_{2}\right\rangle  &  =\left(  -\frac{1}{4}\right)  \sum
_{\mu_{0}}e\left(  \frac{e}{2m}\right)  ^{3}\sum_{\mu_{1},\mathbf{k}%
_{1}\mathbf{,\lambda}_{1}}\sum_{\mu_{2},\mathbf{k}_{2}\mathbf{,\lambda}_{2}%
}\epsilon_{x1}^{2}\epsilon_{x2}^{2}\mathfrak{h}_{1}^{2}\mathfrak{h}_{2}%
^{2}(k_{x2}^{2})\nonumber\\
&  \times\left[  1+\mu_{0}\frac{(\mu_{1}\omega_{1}+\mu_{2}\omega_{2})}%
{\omega_{0}}\right]  \frac{(-i\mu_{0}\omega_{0})}{C_{1}C_{(-1)}D_{1+2}}%
\int_{0}^{\tau}dtR_{0}R_{(-1)}R_{(-2)} \label{W22}%
\end{align}
Now because the denominator involves $C_{1}C_{(-1)}=[-(\mu_{1}\omega_{1}%
)^{2}+\omega_{0}^{2}-i\Gamma(\mu_{1}\omega_{1})^{3}][-(-\mu_{1}\omega_{1}%
)^{2}+\omega_{0}^{2}-i\Gamma(-\mu_{1}\omega_{1})^{3}]=(-\omega_{1}^{2}%
+\omega_{0}^{2})^{2}+(\Gamma\omega_{1}^{3})^{2},$ while the constant $\Gamma$
is assumed small, the function in the sum over $\mathbf{k}_{1}$ is sharply
peaked at $\omega_{1}=\omega_{0}.$ \ Then we approximate the sum over
$\mathbf{k}_{1}$ by an integral and set $\omega_{1}=\omega_{0}$ everywhere
except where the combination $\omega_{1}-\omega_{0}$ appears. \ We extend the
lower limit on the integral to $-\infty,$ and so evaluate the sum using Eqs.
(\ref{SI0})\ and (\ref{ang}) as%
\begin{align}
&  \sum_{\mu_{1},\mathbf{k}_{1}\mathbf{,\lambda}_{1}}\epsilon_{x1}%
^{2}\mathfrak{h}_{1}^{2}\left[  1+\mu_{0}\frac{(\mu_{1}\omega_{1}+\mu
_{2}\omega_{2})}{\omega_{0}}\right]  \frac{1}{C_{1}C_{(-1)}}\nonumber\\
&  \approx\sum_{\mu_{1}}\left(  \frac{a}{2\pi}\right)  ^{3}\int_{0}^{\infty
}\frac{d\omega_{1}\omega_{0}^{2}}{c^{3}}\frac{8\pi}{3}\left(  \frac
{8\pi\mathcal{E(}\omega_{0}\mathcal{)}}{a^{3}}\right)  \left[  1+\mu_{0}%
\frac{(\mu_{1}\omega_{0}+\mu_{2}\omega_{2})}{\omega_{0}}\right] \nonumber\\
&  \times\frac{1}{(2\omega_{0})^{2}(-\omega_{1}+\omega_{0})^{2}+(\Gamma
\omega_{0}^{3})^{2}}\nonumber\\
&  \approx\sum_{\mu_{1}}\left(  \frac{1}{\pi}\right)  ^{3}\frac{\omega_{0}%
^{2}}{c^{3}}\frac{8\pi}{3}\left(  \pi\mathcal{E(}\omega_{0}\mathcal{)}\right)
\left[  1+\mu_{0}\frac{(\mu_{1}\omega_{0}+\mu_{2}\omega_{2})}{\omega_{0}%
}\right]  \int_{-\infty}^{\infty}dx\frac{1}{(2\omega_{0})^{2}(x)^{2}%
+(\Gamma\omega_{0}^{3})^{2}}\nonumber\\
&  =\sum_{\mu_{1}}\left(  \frac{1}{\pi}\right)  ^{3}\frac{\omega_{0}^{2}%
}{c^{3}}\frac{8\pi}{3}\left(  \pi\mathcal{E(}\omega_{0}\mathcal{)}\right)
\left[  1+\mu_{0}\frac{(\mu_{1}\omega_{0}+\mu_{2}\omega_{2})}{\omega_{0}%
}\right]  \frac{\pi}{(2\omega_{0})(\Gamma\omega_{0}^{3})}\nonumber\\
&  =\sum_{\mu_{1}}\frac{1}{\Gamma c^{3}}\frac{4}{3}\frac{\mathcal{E(}%
\omega_{0}\mathcal{)}}{\omega_{0}^{2}}\left[  1+\mu_{0}\frac{(\mu_{1}%
\omega_{0}+\mu_{2}\omega_{2})}{\omega_{0}}\right]  .
\end{align}

Now the time integral needed in Eq. (\ref{W22}) is
\begin{align}
\int_{0}^{\tau}dtR_{0}R_{(-1)}R_{(-2)}  &  =\int_{0}^{\tau}dt\exp[(-i\mu
_{0}\omega_{0}+i\mu_{1}\omega_{0}+i\mu_{2}\omega_{2})t]\nonumber\\
&  =\frac{\exp[(-i\mu_{0}\omega_{0}+i\mu_{1}\omega_{0}+i\mu_{2}\omega_{2}%
)\tau]-1}{-i\mu_{0}\omega_{0}+i\mu_{1}\omega_{0}+i\mu_{2}\omega_{2}},
\end{align}
while the constant $D_{1+2}$ is%
\begin{equation}
D_{1+2}=[-(\mu_{1}\omega_{1}+\mu_{2}\omega_{2})^{2}+\omega_{0}^{2}].
\end{equation}
We will have resonance near $\omega_{2}\approx2\omega_{0}$ only if $\mu
_{1}=-\mu_{0}=-\mu_{2}.$ \ Thus performing the sums over $\mu_{1}$ and
$\mu_{2},$ and retaining only the resonant terms, equation (\ref{W22}) becomes%

\begin{align}
\left\langle W_{2}\right\rangle  &  \approx\left(  \frac{1}{4}\right)
\sum_{\mu_{0}}e\left(  \frac{e}{2m}\right)  ^{3}\sum_{\mu_{1}}\sum_{\mu
_{2},\mathbf{k}_{2}\mathbf{,\lambda}_{2}}\epsilon_{x2}^{2}\mathfrak{h}_{2}%
^{2}(k_{x2}^{2})\frac{4\mathcal{E(}\omega_{0}\mathcal{)}}{3\Gamma c^{3}%
\omega_{0}^{2}}\left[  1+\mu_{0}\frac{(\mu_{1}\omega_{0}+\mu_{2}\omega_{2}%
)}{\omega_{0}}\right] \nonumber\\
&  \times\frac{(-i\mu_{0}\omega_{0})}{[-(\mu_{1}\omega_{1}+\mu_{2}\omega
_{2})^{2}+\omega_{0}^{2}]}\frac{1-\exp[(-i\mu_{0}\omega_{0}+i\mu_{1}\omega
_{0}+i\mu_{2}\omega_{2})\tau]}{-i\mu_{0}\omega_{0}+i\mu_{1}\omega_{0}+i\mu
_{2}\omega_{2}}\nonumber\\
&  \approx\left(  \frac{1}{4}\right)  \sum_{\mu_{0}}e\left(  \frac{e}%
{2m}\right)  ^{3}\sum_{\mathbf{k}_{2}\mathbf{,\lambda}_{2}}\epsilon_{x2}%
^{2}\mathfrak{h}_{2}^{2}(k_{x2}^{2})\frac{4\mathcal{E(}\omega_{0}\mathcal{)}%
}{3\Gamma c^{3}\omega_{0}^{2}}\left[  1+\mu_{0}\frac{(-\mu_{0}\omega_{0}%
+\mu_{0}\omega_{2})}{\omega_{0}}\right] \nonumber\\
&  \times\frac{(-i\mu_{0}\omega_{0})}{[-(-\mu_{0}\omega_{0}+\mu_{0}\omega
_{2})^{2}+\omega_{0}^{2}]}\frac{1-\exp[(-i\mu_{0}\omega_{0}-i\mu_{0}\omega
_{0}+i\mu_{0}\omega_{2})\tau]}{-i\mu_{0}\omega_{0}-i\mu_{0}\omega_{0}+i\mu
_{0}\omega_{2}}.
\end{align}
Next, summing over $\mu_{0},$ we reduce the expression to
\begin{align}
\left\langle W_{2}\right\rangle  &  =\left(  \frac{1}{4}\right)  e\left(
\frac{e}{2m}\right)  ^{3}\left\{  \sum_{\mathbf{k}_{2}\mathbf{,\lambda}_{2}%
}\epsilon_{x2}^{2}(k_{x2}^{2})\right\}  \mathfrak{h}_{2}^{2}\frac
{4\mathcal{E(}\omega_{0}\mathcal{)}}{3\Gamma c^{3}\omega_{0}^{2}}\nonumber\\
&  \times\frac{1}{[2\omega_{0}-\omega_{2}]}\frac{2-2\cos[(2\omega_{0}%
-\omega_{2})\tau]}{2\omega_{0}-\omega_{2}}%
\end{align}
since%
\begin{equation}
\lbrack-(-\mu_{0}\omega_{0}+\mu_{0}\omega_{2})^{2}+\omega_{0}^{2}%
]=-(\omega_{0}+\omega_{2})^{2}+\omega_{0}^{2}=\omega_{2}(2\omega_{0}%
-\omega_{2}).
\end{equation}
At this point, we convert the sum to an integral using Eq. (\ref{SI0}) and
note the angular integration in $\mathbf{k}$%

\begin{equation}
\int d\Omega(1-\cos^{2}\theta)\cos^{2}\theta=\int_{0}^{2\pi}d\phi\int_{0}%
^{\pi}d\theta\sin\theta(1-\cos^{2}\theta)\cos^{2}\theta=\frac{8\pi}{15},
\end{equation}
to obtain%
\begin{align}
\left\langle W_{2}\right\rangle  &  =\left(  \frac{1}{4}\right)  e\left(
\frac{e}{2m}\right)  ^{3}\left\{  \left(  \frac{a}{2\pi}\right)  ^{3}\int
d^{3}k\frac{(k^{2}-k_{x}^{2})}{k^{2}}(k_{x2}^{2})\right\}  \left(  \frac
{8\pi\mathcal{E(}\omega_{2}\mathcal{)}}{a^{3}}\right) \nonumber\\
&  \times\frac{4\mathcal{E(}\omega_{0}\mathcal{)}}{3[2e^{2}/(3mc^{3}%
)]c^{3}\omega_{0}^{2}}\frac{2-2\cos[(2\omega_{0}-\omega_{2})\tau]}{\left(
2\omega_{0}-\omega_{2}\right)  ^{2}}\nonumber\\
&  =\frac{e^{2}}{8m^{2}}\left\{  \frac{1}{\pi^{2}}\int_{0}^{\infty}%
\frac{d\omega_{2}\omega_{2}^{4}}{c^{5}}\frac{8\pi}{15}\right\}  \mathcal{E(}%
\omega_{2}\mathcal{)}\frac{\mathcal{E(}\omega_{0}\mathcal{)}}{\omega_{0}^{2}%
}\frac{1-1\cos[(2\omega_{0}-\omega_{2})\tau]}{\left(  2\omega_{0}-\omega
_{2}\right)  ^{2}}.
\end{align}
The expression involves resonance at $\omega_{2}=2\omega_{0}.$ \ Thus we
approximate all the terms in $\omega_{2}$ as $2\omega_{0}$ except those which
involve the resonant combination $\omega_{2}-2\omega_{0}$, and we extend the
lower limit of integration to $-\infty.$ \ The needed integral is given in Eq.
(\ref{GR432}) so that%
\begin{align}
\left\langle W_{2}\right\rangle  &  =\frac{e^{2}}{8m^{2}}\left\{  \frac{1}%
{\pi^{2}}\frac{(2\omega_{0})^{4}}{c^{5}}\frac{8\pi}{15}\right\}
\mathcal{E(}2\omega_{0}\mathcal{)}\left(  \frac{\mathcal{E(}\omega
_{0}\mathcal{)}}{\omega_{0}^{2}}\right)  \pi\tau\nonumber\\
&  =\left(  \frac{16}{15}\right)  \frac{e^{2}(\omega_{0})^{2}}{m^{2}c^{5}%
}\mathcal{E(}2\omega_{0}\mathcal{)E(}\omega_{0}\mathcal{)}\tau. \label{W2av}%
\end{align}

\subsection{Radiation Emitted by the Quadrupole Moment}

In order to obtain the radiation emitted by the quadrupole moment, we consider
the radiation emitted by a point charge moving along the $x$-axis,
$\mathbf{r}_{e}(t)=\widehat{i}x_{e}(t).$ \ The current density is
\begin{equation}
\mathbf{J}_{e}(\mathbf{r},t)=e\mathbf{v}_{e}\delta^{3}[\mathbf{r}%
-\widehat{i}x_{e}(t)]=e\widehat{i}\dot{x}_{e}(t)\{\delta^{3}(\mathbf{r)-}%
x_{e}(t)\partial_{x}\delta^{3}(\mathbf{r})+...\}
\end{equation}
The vector potential due to this current density is%
\begin{align}
\mathbf{A}_{e}(\mathbf{r},t)  &  =\int dt^{\prime}\int d^{3}r^{\prime}%
\frac{\delta(t-t^{\prime}-|\mathbf{r}-\mathbf{r}^{\prime}|/c)}{|\mathbf{r}%
-\mathbf{r}^{\prime}|}\frac{\mathbf{J}_{e}(\mathbf{r}^{\prime},t^{\prime})}%
{c}\nonumber\\
&  =\int dt^{\prime}\int d^{3}r^{\prime}\frac{\delta(t-t^{\prime}%
-|\mathbf{r}-\mathbf{r}^{\prime}|/c)}{c|\mathbf{r}-\mathbf{r}^{\prime}%
|}\{e\widehat{i}\dot{x}_{e}(t^{\prime})\{\delta^{3}(\mathbf{r}^{\prime
}\mathbf{)-}x_{e}(t^{\prime})\partial_{x^{\prime}}\delta^{3}(\mathbf{r}%
^{\prime})+...\}\nonumber\\
&  =\int dt^{\prime}\int d^{3}r^{\prime}\frac{\delta(t-t^{\prime}%
-|\mathbf{r}-\mathbf{r}^{\prime}|/c)}{c|\mathbf{r}-\mathbf{r}^{\prime}%
|}e\widehat{i}\dot{x}_{e}(t^{\prime})\delta^{3}(\mathbf{r}^{\prime}%
\mathbf{)}\nonumber\\
&  -\int dt^{\prime}\int d^{3}r^{\prime}\frac{\delta(t-t^{\prime}%
-|\mathbf{r}-\mathbf{r}^{\prime}|/c)}{c|\mathbf{r}-\mathbf{r}^{\prime}%
|}e\widehat{i}\dot{x}_{e}(t^{\prime})x_{e}(t^{\prime})\partial_{x^{\prime}%
}\delta^{3}(\mathbf{r}^{\prime})+...\nonumber\\
&  =\frac{e\widehat{i}\dot{x}_{e}(t-r/c)}{cr}+\int dt^{\prime}\int
d^{3}r^{\prime}\partial_{x^{\prime}}\left(  \frac{\delta(t-t^{\prime
}-|\mathbf{r}-\mathbf{r}^{\prime}|/c)}{c|\mathbf{r}-\mathbf{r}^{\prime}%
|}\right)  e\widehat{i}\dot{x}_{e}(t^{\prime})x_{e}(t^{\prime})\delta
^{3}(\mathbf{r}^{\prime})\nonumber\\
&  =\frac{e\widehat{i}\dot{x}_{e}(t-r/c)}{cr}-\partial_{x}\left(
\frac{e\widehat{i}\dot{x}_{e}(t-r/c)x_{e}(t-r/c)}{cr}\right)  +...
\label{Aert}%
\end{align}
The vector potential in Eq. (\ref{Aert}) starts with the dipole term and then
includes the quadrupole term. \ For particle motion along a straight line,
there is no magnetic dipole term. \ Now we are interested in only the
\textit{radiation} from the moving charge $e$, and so we go to the radiation
zone. \ Furthermore, we are concerned with only the quadrupole contribution,
which in the radiation zone becomes%
\begin{equation}
\mathbf{A}_{eQ}(\mathbf{r},t)\approx\frac{e\widehat{i}\ddot{x}_{e}%
(t-r/c)[x/(rc)]x_{e}(t-r/c)}{cr}+\frac{e\widehat{i}\dot{x}_{e}(t-r/c)\dot
{x}_{e}(t-r/c)[x/(rc)]}{cr}%
\end{equation}
from the derivative $dr/dx=x/r.$ \ Now we need the magnetic field
$\mathbf{B}_{eQ}=\nabla\times\mathbf{A}_{eQ}.$ \ Then in the radiation zone,
we have%
\begin{equation}
\mathbf{B}_{eQ}\approx-\widehat{r}\times\left\{  e\widehat{i}\left[
\frac{\dddot{x}_{e}(t-r/c)[x/(rc)]x_{e}(t-r/c)}{c^{2}r}+\frac{3\ddot{x}%
_{e}(t-r/c)\dot{x}_{e}(t-r/c)[x/(rc)]}{c^{2}r}\right]  \right\}
\end{equation}
Now treating the $x$-axis as the polar axis, the radiation power emitted per
unit solid angle is $dP/d\Omega=[c/(4\pi)]r^{2}B_{eQ}^{2},$ corresponding to
\begin{align}
\frac{dP}{d\Omega}  &  =\frac{c}{4\pi}\frac{e^{2}}{c^{4}}(\widehat{r}%
\times\widehat{i})^{2}\left\{  \dddot{x}_{e}(t-r/c)[x/(rc)]x_{e}%
(t-r/c)+3\ddot{x}_{e}(t-r/c)\dot{x}_{e}(t-r/c)[x/(rc)]\right\}  ^{2}%
\nonumber\\
&  =\frac{e^{2}}{4\pi c^{5}}\sin^{2}\theta\cos^{2}\theta\left[  \dddot{x}%
_{e}(t-r/c)x_{e}(t-r/c)+3\ddot{x}_{e}(t-r/c)\dot{x}_{e}(t-r/c)\right]  ^{2}%
\end{align}
To obtain the total power radiated, we integrate over all solid angles as%
\begin{equation}
\int d\Omega\sin^{2}\theta\cos^{2}\theta=\int_{0}^{2\pi}d\phi\int_{0}^{\pi
}d\theta\sin\theta\sin^{2}\theta\cos^{2}\theta=\frac{8}{15}\pi
\end{equation}
Then the total power radiated by the quadrupole term is
\begin{equation}
P_{eQ}=\frac{2e^{2}}{15c^{5}}\left[  \dddot{x}_{e}(t-r/c)x_{e}(t-r/c)+3\ddot
{x}_{e}(t-r/c)\dot{x}_{e}(t-r/c)\right]  ^{2}%
\end{equation}

Inserting the expression for $x_{e}(t)$ given in Eq. (\ref{xt}), we have%
\begin{align}
\left\langle P_{eQ}\right\rangle  &  =\left\langle \frac{2e^{2}}{15c^{5}%
}\left[  \left(  \frac{e}{2m}\sum_{\mu_{1},\mathbf{k}_{1}\mathbf{,\lambda}%
_{1}}\epsilon_{x1}\mathfrak{h}_{1}(-i\mu_{1}\omega_{1})^{3}\frac{R_{1}A_{1}%
}{C_{1}}\right)  \left(  \frac{e}{2m}\sum_{\mu_{2},\mathbf{k}_{2}%
\mathbf{,\lambda}_{2}}\epsilon_{x2}\mathfrak{h}_{2}\frac{R_{2}A_{2}}{C_{2}%
}\right)  \right.  \right. \nonumber\\
&  \left.  \left.  +3\left(  \frac{e}{2m}\sum_{\mu_{1},\mathbf{k}%
_{1}\mathbf{,\lambda}_{1}}\epsilon_{x1}\mathfrak{h}_{1}(-i\mu_{1}\omega
_{1})^{2}\frac{R_{1}A_{1}}{C_{1}}\right)  \left(  \frac{e}{2m}\sum_{\mu
_{2},\mathbf{k}_{2}\mathbf{,\lambda}_{2}}\epsilon_{x2}\mathfrak{h}_{2}%
(-i\mu_{2}\omega_{2})\frac{R_{2}A_{2}}{C_{2}}\right)  \right]  ^{2}%
\right\rangle \label{PeQ1}%
\end{align}
Upon squaring the square bracket in Eq. (\ref{PeQ1}), we will be dealing with
three terms. \ The first term from inside the square bracket (corresponding to
the square of the first line in Eq. (\ref{PeQ1})) is%
\begin{align}
T1  &  =\left(  \frac{e}{2m}\sum_{\mu_{1},\mathbf{k}_{1}\mathbf{,\lambda}_{1}%
}\epsilon_{x1}\mathfrak{h}_{1}(-i\mu_{1}\omega_{1})^{3}\frac{R_{1}A_{1}}%
{C_{1}}\right)  \left(  \frac{e}{2m}\sum_{\mu_{2},\mathbf{k}_{2}%
\mathbf{,\lambda}_{2}}\epsilon_{x2}\mathfrak{h}_{2}\frac{R_{2}A_{2}}{C_{2}%
}\right) \nonumber\\
&  \times\left(  \frac{e}{2m}\sum_{\mu_{3},\mathbf{k}_{3}\mathbf{,\lambda}%
_{3}}\epsilon_{x3}\mathfrak{h}_{3}(-i\mu_{3}\omega_{3})^{3}\frac{R_{3}A_{3}%
}{C_{3}}\right)  \left(  \frac{e}{2m}\sum_{\mu_{4},\mathbf{k}_{4}%
\mathbf{,\lambda}_{4}}\epsilon_{x4}\mathfrak{h}_{4}\frac{R_{4}A_{4}}{C_{4}%
}\right) \\
&  =\left(  \frac{e}{2m}\right)  ^{4}\sum_{\mu_{1},\mathbf{k}_{1}%
\mathbf{,\lambda}_{1}}\sum_{\mu_{2},\mathbf{k}_{2}\mathbf{,\lambda}_{2}}%
\sum_{\mu_{3},\mathbf{k}_{3}\mathbf{,\lambda}_{3}}\sum_{\mu_{4},\mathbf{k}%
_{4}\mathbf{,\lambda}_{4}}\epsilon_{x1}\epsilon_{x2}\epsilon_{x3}\epsilon
_{x4}\mathfrak{h}_{1}\mathfrak{h}_{2}\mathfrak{h}_{3}\mathfrak{h}_{4}\\
&  \times(-i\mu_{1}\omega_{1})^{3}(-i\mu_{3}\omega_{3})^{3}\frac{R_{1}%
R_{2}R_{3}R_{4}A_{1}A_{2}A_{3}A_{4}}{C_{1}C_{2}C_{3}C_{4}}%
\end{align}
When we average over the random phases as in Eq. (\ref{4As}), this term
becomes
\begin{equation}
\left\langle T1\right\rangle =\left(  \frac{e}{2m}\right)  ^{4}\sum_{\mu
_{1},\mathbf{k}_{1}\mathbf{,\lambda}_{1}}\sum_{\mu_{2},\mathbf{k}%
_{2}\mathbf{,\lambda}_{2}}\epsilon_{x1}^{2}\epsilon_{x2}^{2}\mathfrak{h}%
_{1}^{2}\mathfrak{h}_{2}^{2}(-i\mu_{1}\omega_{1})^{3}\frac{[2(-i\mu_{2}%
\omega_{2})^{3}+(i\mu_{1}\omega_{1})^{3}]}{C_{1}C_{(-1)}C_{2}C_{(-2)}}%
\end{equation}
since%
\begin{equation}
R_{1}R_{(-1)}=\exp[-i\mu_{1}\omega_{1}]\exp[i\mu_{1}\omega_{1}]=1.
\end{equation}
Now for small charge $e,$ the function $(C_{1}C_{(-1)})^{-1}$ is sharply
peaked at $\omega_{1}=\omega_{0}.$ \ The term involving $(-i\mu_{1}\omega
_{1})^{3}(-i\mu_{2}\omega_{2})^{3}$ is odd in both $\mu_{1}$ and $\mu_{2}$ and
therefore will cancel completely. \ We evaluate the remaining term in the
usual way from Eqs. (\ref{xx})-(\ref{x2av}) as
\begin{align}
\left\langle T1\right\rangle  &  =\left[  \left(  \frac{e}{2m}\right)
^{2}\sum_{\mu_{1},\mathbf{k}_{1}\mathbf{,\lambda}_{1}}\epsilon_{x1}%
^{2}\mathfrak{h}_{1}^{2}\frac{(\mu_{1}\omega_{1})^{6}}{C_{1}C_{(-1)}}\right]
\left[  \left(  \frac{e}{2m}\right)  ^{2}\sum_{\mu_{2},\mathbf{k}%
_{2}\mathbf{,\lambda}_{2}}\epsilon_{x2}^{2}\mathfrak{h}_{2}^{2}\frac{1}%
{C_{2}C_{(-2)}}\right] \nonumber\\
&  =\left[  \left(  \frac{\mathcal{E(}\omega_{0}\mathcal{)}}{m\omega_{0}^{2}%
}\right)  \omega_{0}^{6}\right]  \left(  \frac{\mathcal{E(}\omega
_{0}\mathcal{)}}{m\omega_{0}^{2}}\right)  =\omega_{0}^{6}\left\langle
x^{2}\right\rangle ^{2} \label{TT1}%
\end{align}

The cross term in the square bracket in Eq. (\ref{PeQ1}) is
\begin{align}
T2  &  =2\left(  \frac{e}{2m}\sum_{\mu_{1},\mathbf{k}_{1}\mathbf{,\lambda}%
_{1}}\epsilon_{x1}\mathfrak{h}_{1}(-i\mu_{1}\omega_{1})^{3}\frac{R_{1}A_{1}%
}{C_{1}}\right)  \left(  \frac{e}{2m}\sum_{\mu_{2},\mathbf{k}_{2}%
\mathbf{,\lambda}_{2}}\epsilon_{x2}\mathfrak{h}_{2}\frac{R_{2}A_{2}}{C_{2}%
}\right) \nonumber\\
&  \times3\left(  \frac{e}{2m}\sum_{\mu_{3},\mathbf{k}_{3}\mathbf{,\lambda
}_{3}}\epsilon_{x3}\mathfrak{h}_{3}(-i\mu_{3}\omega_{3})^{2}\frac{R_{3}A_{3}%
}{C_{3}}\right)  \left(  \frac{e}{2m}\sum_{\mu_{4},\mathbf{k}_{4}%
\mathbf{,\lambda}_{4}}\epsilon_{x4}\mathfrak{h}_{4}(-i\mu_{4}\omega_{4}%
)\frac{R_{4}A_{4}}{C_{4}}\right) \\
&  =6\left(  \frac{e}{2m}\right)  ^{4}\sum_{\mu_{1},\mathbf{k}_{1}%
\mathbf{,\lambda}_{1}}\sum_{\mu_{2},\mathbf{k}_{2}\mathbf{,\lambda}_{2}}%
\sum_{\mu_{3},\mathbf{k}_{3}\mathbf{,\lambda}_{3}}\sum_{\mu_{4},\mathbf{k}%
_{4}\mathbf{,\lambda}_{4}}\epsilon_{x1}\epsilon_{x2}\epsilon_{x3}\epsilon
_{x4}\mathfrak{h}_{1}\mathfrak{h}_{2}\mathfrak{h}_{3}\mathfrak{h}_{4}\\
&  \times(-i\mu_{1}\omega_{1})^{3}(-i\mu_{3}\omega_{3})^{2}(-i\mu_{4}%
\omega_{4})\frac{R_{1}R_{2}R_{3}R_{4}A_{1}A_{2}A_{3}A_{4}}{C_{1}C_{2}%
C_{3}C_{4}}%
\end{align}
When we average over the random phases as in Eq. (\ref{4As}), this term
becomes%
\begin{align}
\left\langle T2\right\rangle  &  =6\left(  \frac{e}{2m}\right)  ^{4}\sum
_{\mu_{1},\mathbf{k}_{1}\mathbf{,\lambda}_{1}}\sum_{\mu_{2},\mathbf{k}%
_{2}\mathbf{,\lambda}_{2}}\epsilon_{x1}^{2}\epsilon_{x2}^{2}\mathfrak{h}%
_{1}^{2}\mathfrak{h}_{2}^{2}(-i\mu_{1}\omega_{1})^{3}\nonumber\\
&  \times\frac{\lbrack(-i\mu_{2}\omega_{2})^{2}(i\mu_{2}\omega_{2})+(i\mu
_{1}\omega_{1})^{2}(-i\mu_{2}\omega_{2})+(i\mu_{1}\omega_{1})(-i\mu_{2}%
\omega_{2})^{2}]}{C_{1}C_{(-1)}C_{2}C_{(-2)}}. \label{T2av}%
\end{align}
Only the last term in Eq. (\ref{T2av}) is even both in the number of sums over
$\mu$ and in the number of factors of $\mu,$ and so does not vanish. \ We
evaluate this term in the usual way as%
\begin{align}
\left\langle T2\right\rangle  &  =6\left[  \left(  \frac{e}{2m}\right)
^{2}\sum_{\mu_{1},\mathbf{k}_{1}\mathbf{,\lambda}_{1}}\epsilon_{x1}%
^{2}\mathfrak{h}_{1}^{2}\frac{\omega_{1}^{4}}{C_{1}C_{(-1)}}\right]  \left[
\left(  \frac{e}{2m}\right)  ^{2}\sum_{\mu_{2},\mathbf{k}_{2}\mathbf{,\lambda
}_{2}}\epsilon_{x2}^{2}\mathfrak{h}_{2}^{2}\frac{\omega_{2}^{2}}{C_{2}%
C_{(-2)}}\right] \nonumber\\
&  =6\left[  \left(  \frac{\mathcal{E(}\omega_{0}\mathcal{)}}{m\omega_{0}^{2}%
}\right)  \omega_{0}^{4}\right]  \left[  \left(  \frac{\mathcal{E(}\omega
_{0}\mathcal{)}}{m\omega_{0}^{2}}\right)  \omega_{0}^{2}\right]  =6\omega
_{0}^{6}\left\langle x^{2}\right\rangle ^{2} \label{TT2}%
\end{align}
The third term in the square bracket of Eq. (\ref{PeQ1}) (corresponding to the
square of the second line of the equation) is
\begin{align}
T3  &  =9\left(  \frac{e}{2m}\sum_{\mu_{1},\mathbf{k}_{1}\mathbf{,\lambda}%
_{1}}\epsilon_{x1}\mathfrak{h}_{1}(-i\mu_{1}\omega_{1})^{2}\frac{R_{1}A_{1}%
}{C_{1}}\right)  \left(  \frac{e}{2m}\sum_{\mu_{2},\mathbf{k}_{2}%
\mathbf{,\lambda}_{2}}\epsilon_{x2}\mathfrak{h}_{2}(-i\mu_{2}\omega_{2}%
)\frac{R_{2}A_{2}}{C_{2}}\right) \nonumber\\
&  \times\left(  \frac{e}{2m}\sum_{\mu_{3},\mathbf{k}_{3}\mathbf{,\lambda}%
_{3}}\epsilon_{x3}\mathfrak{h}_{3}(-i\mu_{3}\omega_{3})^{2}\frac{R_{3}A_{3}%
}{C_{3}}\right)  \left(  \frac{e}{2m}\sum_{\mu_{4},\mathbf{k}_{4}%
\mathbf{,\lambda}_{4}}\epsilon_{x4}\mathfrak{h}_{4}(-i\mu_{4}\omega_{4}%
)\frac{R_{4}A_{4}}{C_{4}}\right) \nonumber\\
&  =9\left(  \frac{e}{2m}\right)  ^{4}\sum_{\mu_{1},\mathbf{k}_{1}%
\mathbf{,\lambda}_{1}}\sum_{\mu_{2},\mathbf{k}_{2}\mathbf{,\lambda}_{2}}%
\sum_{\mu_{3},\mathbf{k}_{3}\mathbf{,\lambda}_{3}}\sum_{\mu_{4},\mathbf{k}%
_{4}\mathbf{,\lambda}_{4}}\epsilon_{x1}\epsilon_{x2}\epsilon_{x3}\epsilon
_{x4}\mathfrak{h}_{1}\mathfrak{h}_{2}\mathfrak{h}_{3}\mathfrak{h}%
_{4}\nonumber\\
&  \times(-i\mu_{1}\omega_{1})^{2}(-i\mu_{2}\omega_{2})(-i\mu_{3}\omega
_{3})^{2}(-i\mu_{4}\omega_{4})\frac{R_{1}R_{2}R_{3}R_{4}A_{1}A_{2}A_{3}A_{4}%
}{C_{1}C_{2}C_{3}C_{4}}%
\end{align}
When we average over the random phases as in Eq. (\ref{4As}), this term
becomes%
\begin{align}
\left\langle T3\right\rangle  &  =9\left(  \frac{e}{2m}\right)  ^{4}\sum
_{\mu_{1},\mathbf{k}_{1}\mathbf{,\lambda}_{1}}\sum_{\mu_{2},\mathbf{k}%
_{2}\mathbf{,\lambda}_{2}}\epsilon_{x1}^{2}\epsilon_{x2}^{2}\mathfrak{h}%
_{1}^{2}\mathfrak{h}_{2}^{2}(-i\mu_{1}\omega_{1})^{2}\nonumber\\
&  \times\frac{\lbrack2(i\mu_{1}\omega_{1})(-i\mu_{2}\omega_{2})^{2}(i\mu
_{2}\omega_{2})+(i\mu_{1}\omega_{1})^{2}(-i\mu_{2}\omega_{2})(i\mu_{2}%
\omega_{2})]}{C_{1}C_{(-1)}C_{2}C_{(-2)}}. \label{T3av}%
\end{align}
\ Only the last term in Eq. (\ref{T3av}) has an even number of terms both in
the sums over $\mu$ and in the number of factors of both $\mu_{1}$ and
$\mu_{2}$ and so is nonvanishing. \ We evaluate this term in the usual way as%
\begin{align}
\left\langle T3\right\rangle  &  =9\left[  \left(  \frac{e}{2m}\right)
^{2}\sum_{\mu_{1},\mathbf{k}_{1}\mathbf{,\lambda}_{1}}\epsilon_{x1}%
^{2}\mathfrak{h}_{1}^{2}\frac{\omega_{1}^{4}}{C_{1}C_{(-1)}}\right]  \left[
\left(  \frac{e}{2m}\right)  ^{2}\sum_{\mu_{2},\mathbf{k}_{2}\mathbf{,\lambda
}_{2}}\epsilon_{x2}^{2}\mathfrak{h}_{2}^{2}\frac{\omega_{2}^{2}}{C_{2}%
C_{(-2)}}\right] \nonumber\\
&  =9\left[  \left(  \frac{\mathcal{E(}\omega_{0}\mathcal{)}}{m\omega_{0}^{2}%
}\right)  \omega_{0}^{4}\right]  \left[  \left(  \frac{\mathcal{E(}\omega
_{0}\mathcal{)}}{m\omega_{0}^{2}}\right)  \omega_{0}^{2}\right]  =9\omega
_{0}^{6}\left\langle x^{2}\right\rangle ^{2} \label{TT3}%
\end{align}

Summing the terms in Eqs. (\ref{TT1}), (\ref{TT2}), and (\ref{TT3}), equation
(\ref{PeQ1}) becomes%
\begin{equation}
\left\langle P_{eQ}\right\rangle =\frac{2e^{2}}{15c^{5}}\omega_{0}^{6}\left(
\frac{\mathcal{E(}\omega_{0}\mathcal{)}}{m\omega_{0}^{2}}\right)
^{2}(1+6+9)=\frac{32e^{2}}{15c^{5}}\omega_{0}^{6}\left(  \frac{\mathcal{E(}%
\omega_{0}\mathcal{)}}{m\omega_{0}^{2}}\right)  ^{2}. \label{PeQF}%
\end{equation}
Equating the average energy absorbed at frequency $2\omega_{0}$ in Eq.
(\ref{W2av}) with the average energy emitted at $2\omega_{0}$ from Eq.
(\ref{PeQF}) during the short time interval $\tau$, we have
\begin{equation}
\left(  \frac{16}{15}\right)  \frac{e^{2}(\omega_{0})^{2}}{m^{2}c^{5}%
}\mathcal{E(}2\omega_{0}\mathcal{)E(}\omega_{0}\mathcal{)}\tau.=\frac{32e^{2}%
}{15c^{5}}\omega_{0}^{6}\left(  \frac{\mathcal{E(}\omega_{0}\mathcal{)}%
}{m\omega_{0}^{2}}\right)  ^{2}\tau. \label{Bal2}%
\end{equation}
Energy balance requires
\begin{equation}
\mathcal{E(}2\omega_{0}\mathcal{)}=2\mathcal{E}(\omega_{0}).
\end{equation}
\ This condition states that the spectrum is linear in frequency, which is
precisely the character of the Lorentz-invariant spectrum of zero-point
radiation. \ 

\section{Comments on the Physical Situation}

\subsection{Higher Multipoles and Radiation}

The mechanical harmonic-oscillator scattering system which we have considered
may be regarded as relativistic provided that $\omega_{0}^{2}\left\langle
x^{2}\right\rangle <<c^{2}.$ \ The usual \textit{point}-dipole approximation
for both the mechanical motion and the interaction with radiation corresponds
to the limit $\left\langle x^{2}\right\rangle \rightarrow0$ while
$e^{2}\rightarrow\infty$ in such a way that the dipole moment squared $p^{2}$
is finite $p^{2}=e^{2}\left\langle x^{2}\right\rangle =const.$ \ In the limit
$\left\langle x^{2}\right\rangle \rightarrow0,$ the mechanical system clearly
satisfies the limit of nonrelativistic speed, and all the radiation emission
and absorption takes place at the fundamental frequency; there is no radiation
interaction at the harmonics since all the higher multipole moments above the
dipole moment vanish. \ In the analysis of the present article, we avoid the
limit $\left\langle x^{2}\right\rangle \rightarrow0.$ \ The amplitude of
oscillator motion is required to be so small that the speed of the particle is
nonrelativistic, but the quadrupole moment is non-vanishing. \ The quadrupole
moment involves two factors of length (not just the one factor of length
needed for the dipole moment) and so vanishes in the usual point-dipole limit.
\ Indeed, all the multipole moments above the dipole moment have additional
factors of length and so require a non-zero amplitude of motion in order to
remain non-zero.

\subsection{Role of the Constant c}

The mechanical motion of the oscillator involves no factors of the speed of
light $c$ so long as the speed $v$ of the oscillator is small $v<<c$. \ On the
other hand, the radiation energy emitted and the radiation energy absorbed at
each harmonic involve the same number of factors of $c$, so that the condition
of radiation balance, harmonic-by-harmonic, gives no role for the ratio $v/c.$
\ Thus the radiation at the fundamental frequency involves factors of $c^{-3}$
for both the emitted radiation and the absorbed radiation, as seen in Eq.
(\ref{Bal1}), while the radiation at the first harmonic involves balancing
factors of $c^{-5},$ as seen in Eq. (\ref{Bal2}). Accordingly, as far as an
analysis harmonic-by-harmonic is involved, there is no connection between the
oscillator speed $v$ and the radiation speed $c.$ It is only when we sum the
series for the radiation emission or absorption that we discover that there is
a singularity associated with the radiation interaction when the particle
speed $v$ approaches the speed of light $c.$\cite{Burko}

\subsection{Adiabatic Invariance}

In the past, it has been shown that the adiabatic invariance of the
\textit{point} harmonic oscillator fits with the adiabatic invariance of
classical electromagnetic zero-point radiation.\cite{B1978c} \ Thus as the
frequency of the oscillator is changed adiabatically, the harmonic oscillator
remains in radiation balance with the zero-point radiation. \ In the earlier
work, it was emphasized that the adiabatic invariance in the presence of
radiation depended crucially upon the absence of any interaction with
radiation harmonics. Based upon the present work, we see that this adiabatic
invariance extends beyond the \textit{point} oscillator out to an oscillator
of finite non-zero amplitude. \ A small oscillator of non-zero excursion has
detailed balance with zero-point radiation at higher harmonics. \ Since the
zero-point radiation is invariant under a $\sigma_{ltU^{-1}}$-scale
transformation\cite{B1989a}, the adiabatic invariance under a change of
oscillator frequency will continue for a harmonic oscillator of small but
non-zero amplitude. \ 

\subsection{Limitations on the Approximation}

The charged particle in a harmonic-oscillator potential which is used in the
calculation of this article has a distinct limitation. \ As described here,
the system involves acceleration-based radiation emission, but does not allow
any velocity-dependent damping proportional to the random radiation which is
present. \ Thus our description does not allow the treatment of random
radiation involving velocity-dependent damping. \ Now thermal radiation has a
preferred inertial frame, and any particle moving relative to this preferred
inertial frame will experience a velocity-dependent damping proportional to
the thermal radiation which is present.\cite{B1969b} \ On the other hand,
zero-point radiation is Lorentz invariant and so involves no
velocity-dependent damping. \ The mathematical description used in the present
article is accurate for zero-point radiation only and does not extend to
thermal radiation at non-zero temperature. \ 

\section{Discussion of the Connection Between Blackbody Radiation and
Relativity}

\subsection{Relativistic Invariance is Strongly Restrictive}

The basic physical ideas involved in the present calculation are not well
known and deserve a broader audience. \ Many physicists are unaware of the
restrictive nature of the requirement that a system should be relativistic.
\ The first three conservation laws of dynamics, associated with symmetries
under space translations, time translations, and rotations, involve
conservation of linear momentum, energy, and angular momentum; these
conservation laws appear in both nonrelativistic and relativistic theories.
\ However, the fourth conservation law associated \ with Galilean symmetry or
relativistic symmetry is quite different between nonrelativistic and
relativistic systems.\cite{B2009a} \ The conservation law for Galilean
invariance of nonrelativistic dynamics essentially repeats the information of
the law of conservation of linear momentum and so gives no restrictions on
allowed nonrelativistic systems. \ On the other hand, the conservation law
associated with Lorentz invariance has profound limitations on which systems
are relativistic. For example, the no-interaction theorem of Currie, Jordan,
and Sudarshan\cite{Currie} states that any relativistic interaction between
particles beyond point interactions requires the introduction of a field
theory, and relativistic field theories are restricted still further.

\subsection{Simplest Equilibrium Radiation Spectra}

Within classical physics, there are two spectra for electromagnetic radiation
which take particularly simple forms. \ One of these is the Rayleigh-Jeans
spectrum which associates the same energy (taken as $k_{B}T)$ with every
radiation normal mode. \ This spectrum involves one parameter, the energy
$k_{B}T$ per normal mode, and does not distinguish any length or any
frequency. \ This spectrum is associated with nonrelativistic physics and in
particular with the equipartition theorem of nonrelativistic classical
statistical mechanics.

The second simple radiation spectrum is that of classical electromagnetic
zero-point radiation which has an energy linear in the frequency (energy taken
as $\hbar\omega/2)$ with every radiation normal mode. \ Again, this spectrum
involves one parameter $\hbar$ with units corresponding to an angular momentum
or to an energy $\times$ time, and does not distinguish any length or
frequency. \ The spectrum is\ associated with relativity; zero-point radiation
is the unique (up to a multiplicative constant) Lorentz-invariant spectrum of
random classical radiation, and it takes the same form in every inertial frame.

It is striking that the Planck blackbody radiation spectrum including
zero-point radiation is the smoothest possible interpolation between the
Rayleigh-Jeans spectrum at low frequencies and the zero-point radiation
spectrum at high frequencies.\cite{B2003c}

\subsection{Scatterers for Classical Radiation Equilibrium}

Electromagnetic radiation can not bring itself to equilibrium. \ Rather, there
must be some mechanical scattering system, some \textquotedblleft
black\textquotedblright\ particle (a particle which scatters radiation toward
the equilibrium spectrum), which enforces the radiation equilibrium. \ Since
radiation equilibrium is determined by a mechanical system, we certainly
expect that the nature of the scattering system will influence the radiation
equilibrium spectrum. \ Indeed for classical mechanical systems,
nonrelativistic scatterers leave the Rayleigh-Jeans spectrum unchanged, while
relativistic scatterers leave the zero-point spectrum unchanged.

The simplest scattering system which allows a transition corresponding to that
found for the Planck spectrum (including zero-point radiation) between the
Rayleigh-Jeans spectrum at low frequency and the zero-point spectrum at high
frequency is that of a charged particle of charge $e$ and mass $m$ in a
Coulomb potential where the ratio $mc^{2}/k_{B}T$ provides the mechanical
transition parameter matching the radiation transition parameter $\hbar
\omega/k_{B}T.$ \ We notice that for the Coulomb potential, high mechanical
mass $m$ is associated with high frequency $\omega$. \ The Coulomb potential
is one of the few mechanical systems where large mass $m$ is associated with
high frequency, and small mass $m$ is associated with low frequency.

\subsection{Harmonic Oscillator Scatterers}

Although hydrogen-like scatterers involving a Coulomb potential are the
relativistic systems which are expected to scatter classical electromagnetic
radiation toward the Planck spectrum with zero-point radiation, it seems
exceedingly difficult to work with the Coulomb potential as a scatterer. \ On
the other hand, it is vastly easier to treat a charged harmonic oscillator as
a scattering system for electromagnetic radiation. \ Indeed, at the end of the
19th century, Planck calculated the behavior of a charged harmonic oscillator
taken in the point-size limit when bathed in random classical electromagnetic
radiation. \ Although Planck had initially hoped that the oscillators would
serve as \textquotedblleft black\textquotedblright\ particles and determine
the spectrum of electromagnetic radiation, it became clear that small
(point-limit) harmonic oscillator systems simply acquired an energy which
matched the energy of the radiation normal modes at the oscillator frequency.
\ A point harmonic dipole oscillator did not determine the equilibrium
radiation spectrum. \ Acting as a scatterer, the oscillator may change the
angular distribution of the radiation, but the point oscillator did not change
the frequency spectrum of the random electromagnetic radiation. \ Planck
subsequently turned to statistical mechanics for the harmonic oscillator in an
attempt to understand equilibrium for the electromagnetic radiation. \ And the
application of the nonrelativistic equipartition theorem to an oscillator
scatterer is still used in physics textbooks as a way of obtaining the
Rayleigh-Jeans radiation spectrum.\cite{err} \ 

\subsection{Extensions for Harmonic Oscillator Scatterers}

We wish to avoid the use of statistical mechanics in our exploration of
radiation equilibrium, but we would like to use the calculational simplicity
of harmonic oscillator systems. \ Now point oscillator systems are
unsatisfactory because they interact with random radiation at a single
frequency. \ However, there are two natural extensions of the point-limit
harmonic oscillator system which will bring the system into contact with the
full radiation spectrum. \ One involves the introduction of a small nonlinear
term in the harmonic oscillator potential so as to introduce higher harmonics
in the oscillator mechanical motion. \ The second possible modification is the
consideration of all the radiation harmonics associated with a charged
particle in a purely harmonic potential but with an amplitude of finite,
non-zero excursion. \ 

\subsubsection{Nonlinear-Oscillator Scatterer}

The radiation scattering due to an oscillator with a small nonlinear term was
considered in 1976.\cite{B1976a} The introduction of a small nonlinear term in
the harmonic-oscillator potential leads to mechanical motion which involves
harmonics of the basic oscillator motion so that the particle displacement
becomes $x(t)=a_{1}\cos(\omega_{0}t+\phi_{1})+a_{2}\cos(2\omega_{0}t+\phi
_{2})+...$. \ The ratio $a_{n}/a_{1}$ of the amplitudes $a_{n}$ compared to
the initial harmonic-oscillator amplitude $a_{1}$ depends upon the arbitrary
strength of the nonlinear term in the potential. \ Although radiation emission
and absorption are still treated in the dipole approximation, the presence of
the harmonics in the mechanical motion brings the oscillator into contact with
not only the radiation at the fundamental oscillator frequency $\omega_{0}$
but also the radiation at the multiples $n\omega_{0}$ of the fundamental
frequency $\omega_{0}.$ \ Although the ratios $a_{n}/a_{1}$ may be arbitrary,
it turns out that the radiation spectrum which has the same energy per normal
mode at every frequency (the Rayleigh-Jeans spectrum) remains unchanged by
scattering from this system. \ Indeed, there have been several calculation,
going back to van Vleck's work of 1924 showing that nonrelativistic nonlinear
mechanical scattering systems treated in the dipole limit for their radiation
interaction leave the Rayleigh-Jeans spectrum invariant.\cite{vanVleck}

\subsubsection{Finite-Amplitude Harmonic Oscillator}

The second possible extension of the small harmonic oscillator scatterer is
what is treated in the calculations of the present article. \ We consider not
a change in the harmonic oscillator mechanical potential but rather a
calculation of the radiation emitted and absorbed at the radiation harmonics
of the fundamental oscillator frequency. \ The oscillator potential remains
unchanged as a harmonic oscillator potential and the free oscillator motion
$x(t)=a_{1}\cos(\omega_{0}t+\phi_{1})$ remains unchanged. \ However,
relativistic classical electrodynamics involves radiation at all the harmonics
for any finite non-zero amplitude of oscillation. \ Thus the relativistic
aspects enter not through the mechanical oscillator motion but through the
radiation theory. \ Our analysis calculates the radiation energy balance for
the second harmonic corresponding to quadrupole radiation and shows that the
spectrum which remains unchanged is the zero-point radiation spectrum. \ This
is the first classical scattering calculation showing explicitly that a
relativistic scattering system indeed leaves the relativistically-invariant
zero-point radiation spectrum unchanged. \ 

\section{Acknowledgement}

The present calculation was prompted by the work of Professor Daniel C. Cole
and of Dr. Wayne Cheng-Wei\ Huang and Professor Herman Batelaan showing the
influence of radiation harmonics back on mechanical systems' motions at the
fundamental frequency. \ I wish to thank Professor Michael C. Boyer for
helpful suggestions regarding the presentation of the calculational results.

(Revised July 6, 2018)


\begin{thebibliography}{99}
\bibitem {note}The compact notations are similar but not identical to those
used in ref. 10.
\end{thebibliography}
\end{document}